\documentclass[12pt]{article}
\pdfoutput=1

\usepackage[utf8]{inputenc}
\usepackage[left=2.55cm, right=2.55cm, top=2.55cm, bottom=2.55cm]{geometry}
\usepackage{amsmath,amssymb,amsbsy}
\usepackage{slashed}
\usepackage{xcolor}
\usepackage{graphicx}
\usepackage{url}
\usepackage{cancel}
\usepackage{cite}
\usepackage[colorlinks=true,allcolors=darkpurple,pdfborder={0 0 0},linktocpage=false]{hyperref}
\usepackage{tabularx,booktabs}
\usepackage{multicol}
\usepackage{units}
\usepackage{xspace}
\usepackage[labelfont=bf]{caption}
\usepackage[section]{placeins}
\usepackage{enumitem}
\usepackage{ulem}

\definecolor{darkred}{rgb}{0.6,0,0}
\definecolor{darkpurple}{rgb}{0.5,0,0.5}

\def\hc{\text{h.c.}}

\def\z2{$\mathbb{Z}_2$}

\def\one{$\mathbf{1}$}
\def\two{$\mathbf{2}$}

\def\GCD{\text{GCD}}
\def\I{\text{I}}
\def\II{\text{II}}
\def\III{\text{III}}
\def\IV{\text{IV}}
\def\V{\text{V}}

\definecolor{avblue}{rgb}{0.0, 0.0, 0.8}
\definecolor{asparagus}{rgb}{0.53, 0.66, 0.42}

\newcommand{\AddrCINVESTAV}{%
  Departamento de F\'isica, Centro de Investigaci\'on y de Estudios Avanzados del Instituto Polit\'ecnico Nacional, Apdo. Postal 14-740, 07000 M\'exico D.F., M\'exico}

\newcommand{\AddrIFIC}{%
  Instituto de F\'{i}sica Corpuscular, CSIC-Universitat de Val\`{e}ncia, 46980 Paterna, Spain}

\newcommand{\AddrFISTEO}{%
  Departament de F\'{\i}sica Te\`{o}rica, Universitat de Val\`{e}ncia, 46100 Burjassot, Spain}

\begin{document}

\vspace*{-2cm}
\begin{flushright}
IFIC/23-01 \\
\vspace*{2mm}
\end{flushright}

\begin{center}
\vspace*{15mm}

\vspace{1cm}
{\Large \bf 
Ultraviolet extensions of the Scotogenic model
} \\
\vspace{1cm}

{\bf Diego Portillo-S\'anchez$^{\text{a}}$, Pablo Escribano$^{\text{b}}$, Avelino Vicente$^{\text{b,c}}$}

 \vspace*{.5cm} 
 $^{(\text{a})}$ \AddrCINVESTAV \\\vspace*{.2cm} 
 $^{(\text{b})}$ \AddrIFIC \\\vspace*{.2cm} 
 $^{(\text{c})}$ \AddrFISTEO

 \vspace*{.3cm}
\href{mailto:diego.portillo@cinvestav.mx}{diego.portillo@cinvestav.mx},
\href{mailto:pablo.escribano@ific.uv.es}{pablo.escribano@ific.uv.es},
\href{mailto:avelino.vicente@ific.uv.es}{avelino.vicente@ific.uv.es}
\end{center}

\vspace*{10mm}
\begin{abstract}\noindent\normalsize
The Scotogenic model is a popular scenario that induces radiative
Majorana neutrino masses and includes a weakly-interacting dark matter
candidate. We classify all possible ultraviolet extensions of the
Scotogenic model in which (i) the \textit{dark} \z2 parity emerges at
low energies after the spontaneous breaking of a global $\rm U(1)_L$
lepton number symmetry, and (ii) the low-energy effective theory
contains a naturally small lepton number breaking parameter,
suppressed by the mass of a heavy mediator integrated out at
tree-level. We find $50$ such models and discuss two of them in detail
to illustrate our setup. We also discuss some general aspects of the
phenomenology of the models in our classification, exploring possible
lepton flavor violating signals, collider signatures and implications
for dark matter. The phenomenological prospects of these scenarios are
very rich due to the presence of additional scalar states, including a
massless Goldstone boson.
\end{abstract}

\section{Introduction}
\label{sec:intro}

The Scotogenic model~\cite{Ma:2006km} is a popular extension of the
Standard Model (SM) that addresses two of the currently most important
open questions in physics: the origin of neutrino masses and the
nature of the dark matter (DM) of the Universe. Its popularity stems
from its simplicity. The model extends the SM particle content with
three singlet fermions, $N_{1,2,3}$, and a scalar doublet, $\eta$, all
odd under a new \z2 symmetry under which the SM fields are even. These
ingredients suffice to induce Majorana neutrino masses at the 1-loop
level and provide a viable DM candidate, namely the lightest \z2-odd
state.

Radiative neutrino mass
models~\cite{Zee:1980ai,Cheng:1980qt,Zee:1985id,Babu:1988ki} provide a
natural suppression for neutrino masses with loop factors. This is one
of the main motivations in favor of this class of
models~\cite{Cai:2017jrq}. In addition, further suppression is
introduced in some models by assuming an approximate lepton number
symmetry, broken in a small amount by the presence of a Lagrangian
term with a suppressed coefficient. This is the case of the Scotogenic
model, that requires a small $\lambda_5 \ll 1$ quartic parameter to
obtain the correct size for neutrino masses with sizable Yukawa
couplings. While this is technically valid, and natural in the sense
of 't Hooft~\cite{tHooft:1979rat}, it also calls for an extension that
explains the smallness of the $\lambda_5$ parameter, possibly relating
it to the breaking of lepton number.

In this work we consider ultraviolet (UV) extensions of the Scotogenic
model that provide a natural explanation for the smallness of the
$\lambda_5$ parameter and in which the \z2 parity of the model emerges
at low energies from a spontaneously broken global $\rm U(1)$ lepton
number symmetry. This endeavor was initiated
in~\cite{Escribano:2021ymx}, where a specific UV model with these
properties was proposed. Here we go beyond specific realizations and
classify all possible models with these features in which a low-energy
Scotogenic model is obtained after integrating out a heavy field at
tree-level. Besides one or several massive scalars, the particle
spectrum of the theory will contain a massless Goldstone boson, the
\textit{majoron}~\cite{Chikashige:1980ui,Gelmini:1980re,Schechter:1981cv,Aulakh:1982yn},
induced by the spontaneous breaking of lepton number. These new states
are not present in the original Scotogenic model and lead to novel
phenomenological predictions that allow one to probe our setup.

The rest of the manuscript is organized as follows. First, we set our
notation and conventions in Sec.~\ref{sec:scot}, where the Scotogenic
model is introduced. A general classification of all possible UV
extensions of the Scotogenic model satisfying the requirements
explained above is given in Sec.~\ref{sec:clasif}. Two selected
example models will be presented in detail in Secs.~\ref{sec:model1}
and \ref{sec:model2}. Some general aspects of the phenomenology of
this class of models are discussed in Sec.~\ref{sec:pheno}. Finally,
we summarize our results and conclude in
Sec.~\ref{sec:conclusions}. Additional information can be found in
Appendix~\ref{sec:app}, where we discuss scenarios with an accidental
\z2 symmetry.

\section{The Scotogenic model}
\label{sec:scot}

Before we discuss specific UV realizations of our setup, let us
introduce our conventions for the Scotogenic model. The particle
content of the Scotogenic model~\cite{Ma:2006km} includes, besides the
usual SM fields, three generations of right-handed fermions $N$,
transforming as $\left( \mathbf{1},0\right)$ under $\rm \left( \rm
SU(2)_L , U(1)_Y \right)$, and one scalar $\eta$, transforming as
$\left( \mathbf{2},1/2\right)$. We also impose the conservation of an
ad-hoc \z2 symmetry, under which $\eta$ and $N$ are odd while the rest
of the fields in the model are even. The lepton and scalar particle
content of the model is shown in
Table~\ref{tab:ParticleContent}.~\footnote{We follow the conventions
for the Scotogenic model used in~\cite{Escribano:2020iqq}.}

{
\renewcommand{\arraystretch}{1.4}
\begin{table}[tb]
  \centering
  \begin{tabular}{|c|c||ccc|c|}
      \hline
      \textbf{Field} & \textbf{Generations} & \textbf{$\rm SU(3)_c$} & \textbf{$\rm SU(2)_L$} & \textbf{$\rm U(1)_Y$} & $\boldsymbol{\mathbb{Z}_2}$ \\
      \hline
      $\ell_L$      &  3  &  \one  &  \two  &  -1/2  &  $+$   \\
      $e_R$    &  3  &  \one  &  \one  &  -1    &  $+$   \\
      $N$      &  3  &  \one  &  \one  &  0     &  $-$  \\
      \hline
      $H$      &  1  &  \one  &  \two  &  1/2   &  $+$   \\
      $\eta$   &  1  &  \one  &  \two  &  1/2   &  $-$  \\
      \hline
  \end{tabular}
  \caption{\label{tab:ParticleContent}
    Lepton and scalar particle content and representations under the gauge and discrete symmetries in the Scotogenic model. $\ell_L$ and $e_R$ are the SM left- and right-handed leptons, respectively, and $H$ is the SM Higgs doublet.}
\end{table}
}

\begin{figure}[tb!]
  \centering
  \includegraphics[width=0.5\linewidth]{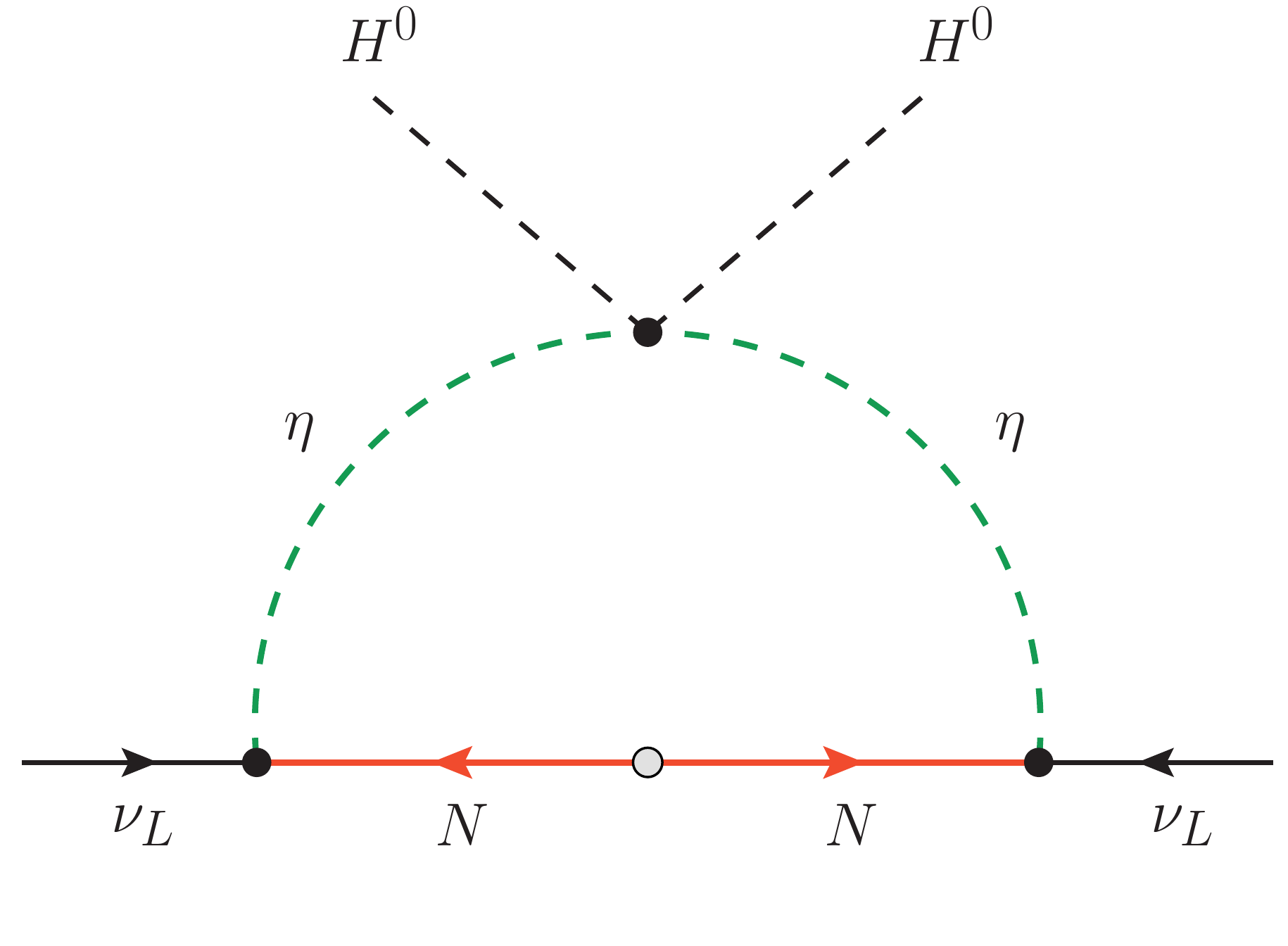}
  \caption{Neutrino mass generation in the Scotogenic model. This Feynman diagram shows the relevant gauge eigenstates involved in the 1-loop contribution to neutrino masses.
    \label{fig:scot}
    }
\end{figure}

The model contains two scalar doublets, the usual Higgs doublet $H$ and
the new doublet $\eta$, only distinguished by their \z2 charges. They
can be decomposed in terms of their $\rm SU(2)_L$ components as
\begin{equation}
  H = \begin{pmatrix}
    H^+ \\
    H^0 
  \end{pmatrix} \, , \quad \eta = \begin{pmatrix}
    \eta^+ \\
    \eta^0 
  \end{pmatrix} \, .
\end{equation}
Once specified the particle content and symmetries of the model we can
write down the Lagrangian. The Lagrangian of the model contains the
terms
\begin{equation}
  \mathcal{L}_{\rm Y} = y \, \overline{N} \, \widetilde{\eta}^\dagger \, \ell_L + \frac{1}{2} M_N \,  \overline{N}^c N + \hc \, , \label{Yscot}
\end{equation}
where $y$ is a general complex $3 \times 3$ matrix and $M_N$ is a
symmetric $3 \times 3$ mass matrix. The scalar potential of the model
is given by
\begin{equation}
\begin{split}
\mathcal{V}_{\rm UV} &= m_H^2 H^{\dagger} H+ m_{\eta}^2\eta^{\dagger}\eta +\frac{\lambda_1}{2}(H^{\dagger}H)^2+\frac{\lambda_2}{2} (\eta^{\dagger} \eta)^2 \\
&+ \lambda_3 (H^{\dagger}H)(\eta^{\dagger}\eta)+\lambda_4 (H^{\dagger}\eta)(\eta^{\dagger}H) +\left[\frac{\lambda_5}{2} (H^{\dagger}\eta)^2 + \hc \right] \, .
\end{split}\label{Vscot}
\end{equation}
Here $m_H^2$ and $m_\eta^2$ are parameters with dimensions of
mass$^2$. We assume that the minimization of the scalar potential
leads to a vacuum defined by
\begin{equation}
  \langle H^0 \rangle = \frac{v_H}{\sqrt{2}} \, , \quad \langle \eta^0 \rangle = 0 \, .
\end{equation}
This vacuum configuration breaks the electroweak symmetry in the usual
way but preserves the \z2 symmetry of the model. As a consequence of
this, the lightest \z2-odd state (either $N_1$ or $\eta^0$) is
completely stable and can play the role of the DM of the
Universe. Furthermore, neutrinos acquire non-zero Majorana masses at
the 1-loop level, as shown in Fig.~\ref{fig:scot}. The resulting $3
\times 3$ neutrino mass matrix is given by
\begin{equation} \label{eq:numass}
  (m_\nu)_{\alpha \beta} = \frac{\lambda_5 \, v_H^2}{32 \pi^2} \sum_n \frac{y_{n \alpha} \, y_{n \beta}}{M_{N_n}} \left[\frac{M_{N_n}^2}{m_0^2 - M_{N_n}^2} + \frac{M_{N_n}^4}{\left(m_0^2 - M_{N_n}^2\right)^2} \, \log\frac{M^2_{N_n}}{m^2_0}\right] \, ,
\end{equation}
where $m_0^2 = m_\eta^2 + \left( \lambda_3 + \lambda_4 \right) \,
v_H^2/2$ and $M_{N_n}$ are the diagonal elements of the $M_N$
matrix. One can easily estimate that in order to obtain neutrino
masses of the order of $0.1$ eV with Scotogenic states in the TeV
scale and Yukawas of order $1$, $\lambda_5$ must be of order $\sim
10^{-10}$. The smallness of this parameter is protected by lepton
number, and thus is technically
natural~\cite{tHooft:1979rat}. However, it is not explained in the
context of the Scotogenic model.

\section{Ultraviolet extensions of the Scotogenic model}
\label{sec:clasif}

\subsection{General considerations}
\label{subsec:general}

The Scotogenic model has two features that call for a refinement,
namely, the origin of the \z2 symmetry and $\lambda_5 \ll 1$. Although
these features do not pose any theoretical problem, they can be
regarded as ad-hoc ingredients in an otherwise very natural
framework. We are thus interested in an UV extension of the Scotogenic
model that provides an explanation for them. More specifically, we
want to classify all possible UV scenarios that lead to the Scotogenic
model at low energies after integrating out a heavy scalar field $S$,
with $m_S \gg v_H$, and satisfy the following two requirements:
\begin{itemize}
\item[{\bf(A)}] The Scotogenic \z2 is obtained as a remnant after the
  spontaneous breaking of a $\rm U(1)_L$ lepton number symmetry by the
  VEV of one or several singlet scalar fields $\sigma$:
  \begin{equation*}
    {\rm U(1)_L} \, \xrightarrow{\quad \langle \sigma \rangle \quad} \, \mathbb{Z}_2
  \end{equation*}
\item[{\bf(B)}] The $(H^\dagger \eta)^2$ operator is forbidden in the UV
  theory due to $\rm U(1)_L$ conservation, but an operator of the form
  $(H^\dagger \eta)^2 \sigma^n$, with $n \geq 1$, is generated after
  integrating out $S$. After the singlets get VEVs and $\rm U(1)_L$ is
  spontaneously broken, this will induce an effective $\lambda_5$
  coupling, which will be naturally suppressed by the large $m_S$
  energy scale.
\end{itemize}
In this work we will concentrate on global $\rm U(1)_L$ lepton number
symmetries, tree-level completions of the $\lambda_5$ operator and UV
models with one or two $\sigma$ singlets. Gauged versions of the
lepton number symmetry, higher-order completions and models with
additional singlets are left for future work.

The models we are looking for induce neutrino masses \textit{\`a la
  Scotogenic}, with variations of the neutrino mass diagram in
Fig.~\ref{fig:scot}. This diagram has an internal scalar line (with
$\eta^0$) and an internal fermion line (with $N$). The analogous
diagrams in the UV extended models will include the heavy scalar $S$
in the loop and one or several external legs with $\sigma$ singlets
(or $\sigma$ insertions, for short). After these considerations, there
are two classes of models that can be already discarded:
\begin{itemize}
\item Models without $\sigma$ insertions in the scalar line. These
  models can be discarded because the $(H^\dagger \eta)^2$ operator
  would be allowed in the UV theory. This would preclude an
  explanation of $\lambda_5 \ll 1$. In addition, $\eta$ would acquire a VEV.
\item Models without $\sigma$ insertions in the fermion line. The $\rm
  U(1)_L$ charge of the $N$ singlet fermions must necessarily vanish
  if the $\sigma \overline{N}^c N$ operator is absent and their
  Majorana masses are explicitly introduced in the
  Lagrangian. However, in this case $N$ will be even under the \z2
  symmetry obtained after spontaneous $\rm U(1)_L$ breaking. This
  scenario does not correspond to the Scotogenic model. Nevertheless,
  an additional accidental \z2 symmetry may appear, as explained in
  Appendix~\ref{sec:app}.
\end{itemize}

\begin{table}[tb]
\centering
\begin{tabular}{ccc}\hline\hline
{\bf Topology} & {\bf Diagram} & {\bf Required operators} \\\hline\hline\\[-1.5ex]
\vspace*{0.2cm}
\begin{tabular}{l}
I
\end{tabular}&\begin{tabular}{l}
\includegraphics[scale=.2]{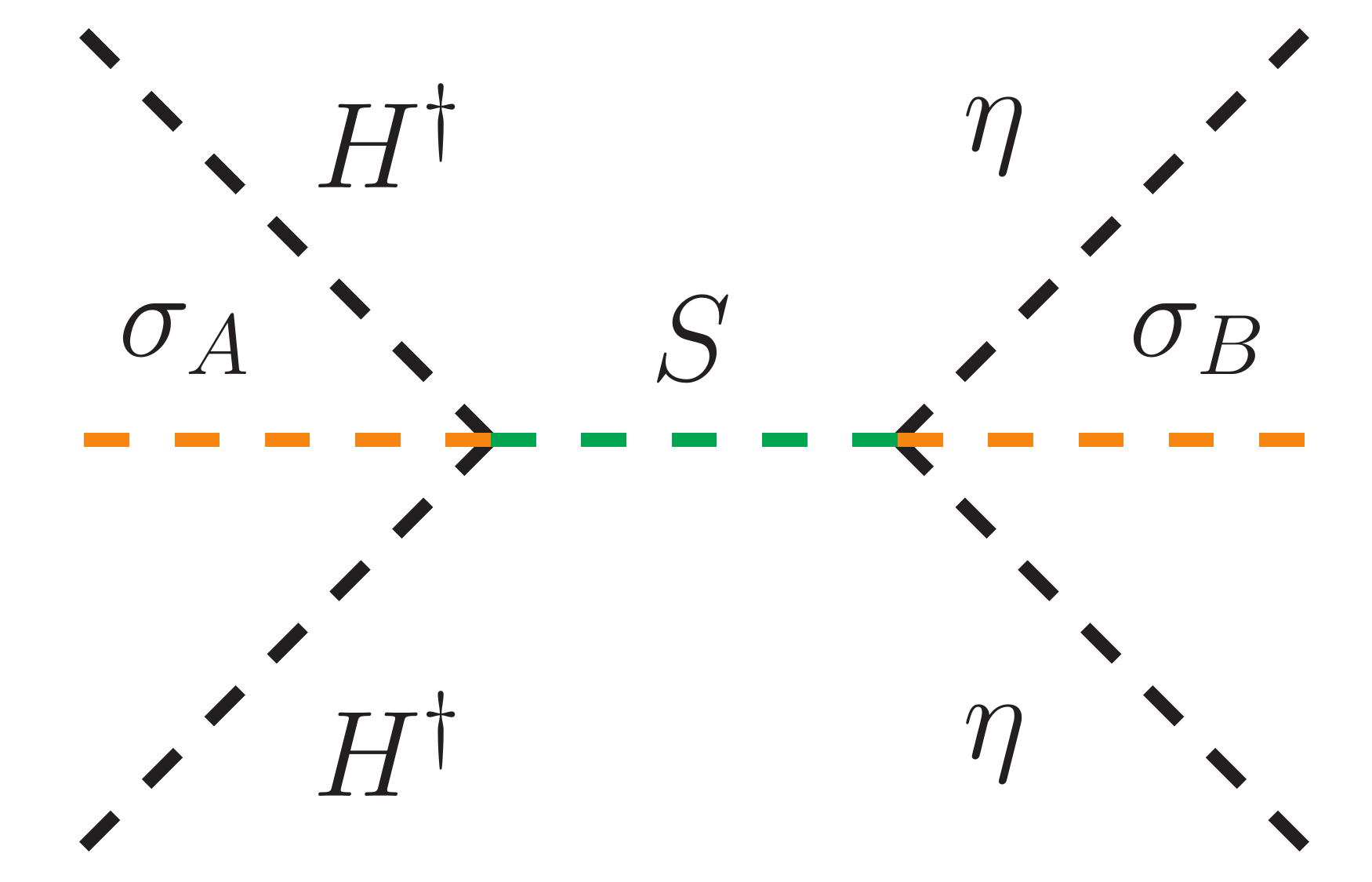}
\end{tabular}&\begin{tabular}{l}
($\sigma_A H^{\dagger}S \tilde{H}$),($\sigma_B \tilde{\eta}^{\dagger}S^{\dagger}\eta$)
\end{tabular}\\
\vspace*{0.2cm}
\begin{tabular}{l}
II
\end{tabular}&\begin{tabular}{l}
\includegraphics[scale=.2]{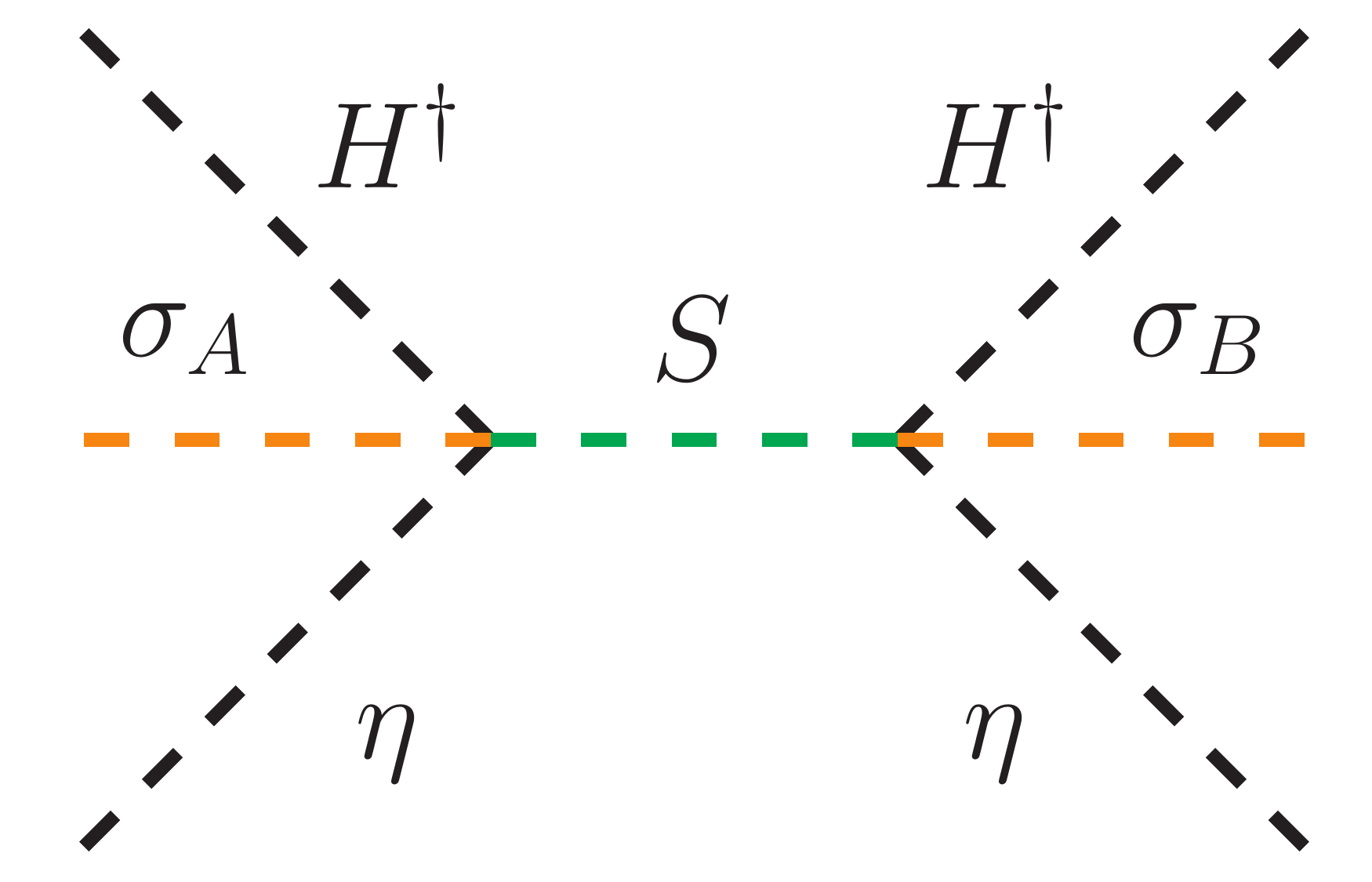}
\end{tabular}&
\begin{tabular}{l}
($\sigma_A H^{\dagger} S \eta$),($\sigma_B H^{\dagger}S^{\dagger}\eta$)
\end{tabular}\\
\vspace*{0.2cm}
\begin{tabular}{l}
III
\end{tabular}&\begin{tabular}{l}
\includegraphics[scale=.2]{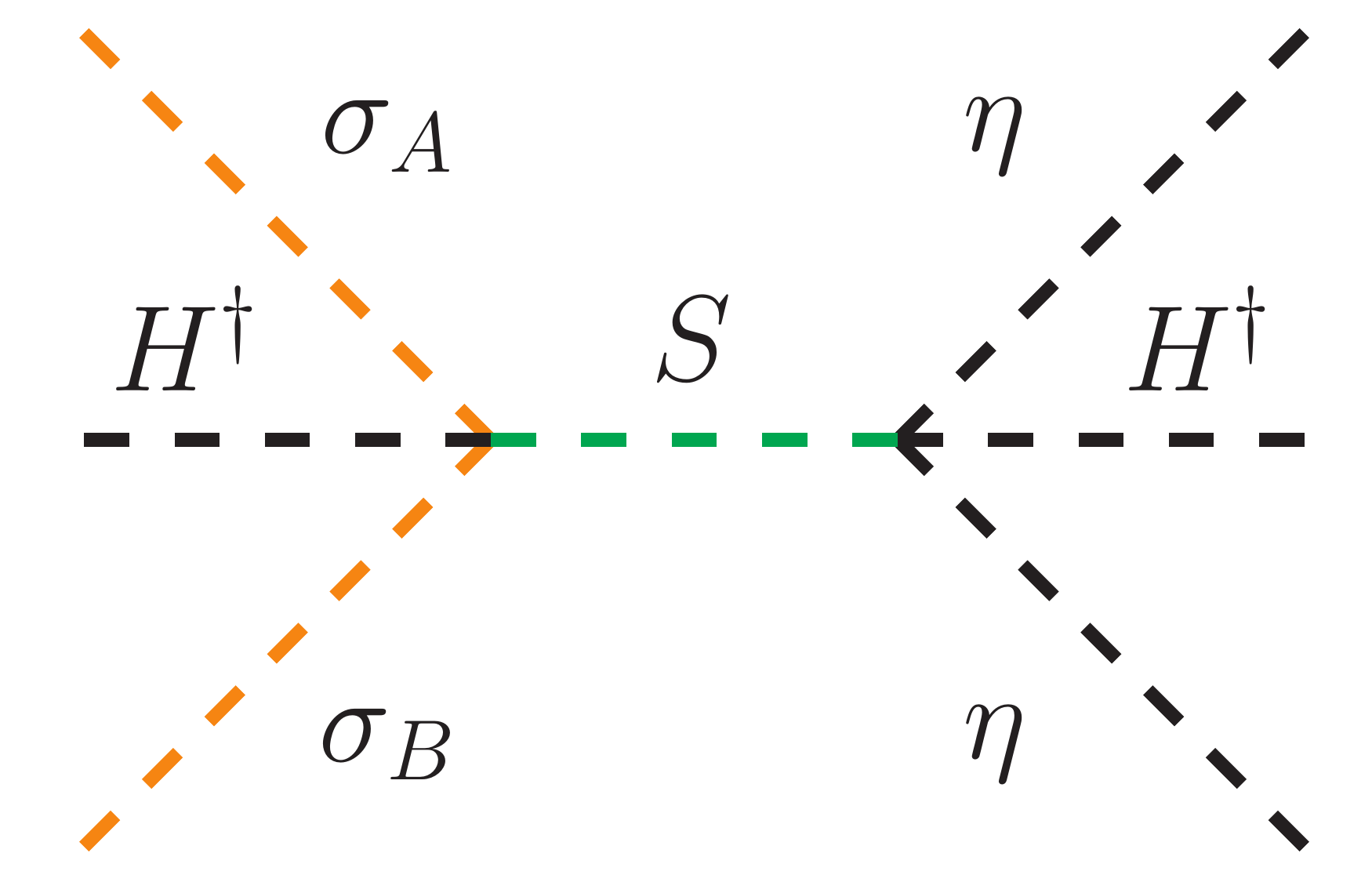}
\end{tabular}&\begin{tabular}{l}
($\sigma_A \sigma_B H^{\dagger} S$),($H^{\dagger}\eta S^{\dagger}\eta$)
\end{tabular}\\
\vspace*{0.1cm}
\begin{tabular}{l}
IV
\end{tabular}&\begin{tabular}{l}
\includegraphics[scale=.2]{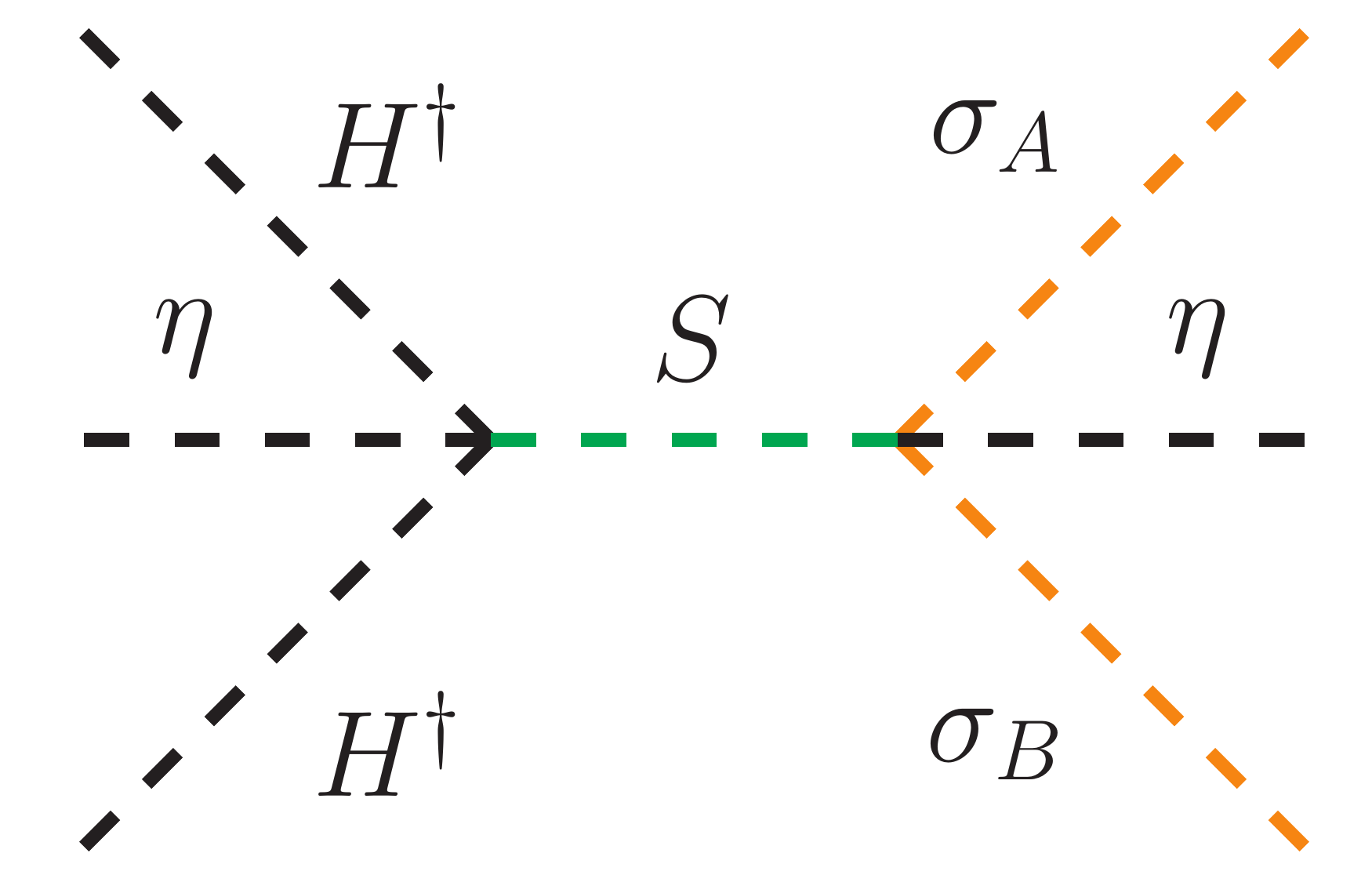}
\end{tabular}&\begin{tabular}{l}
($H^{\dagger} S H^{\dagger} \eta$),($\sigma_A \sigma_B S^{\dagger}\eta$)
\end{tabular}\\
\vspace*{0.2cm}
\begin{tabular}{l}
V
\end{tabular}&\begin{tabular}{l}
\includegraphics[scale=.2]{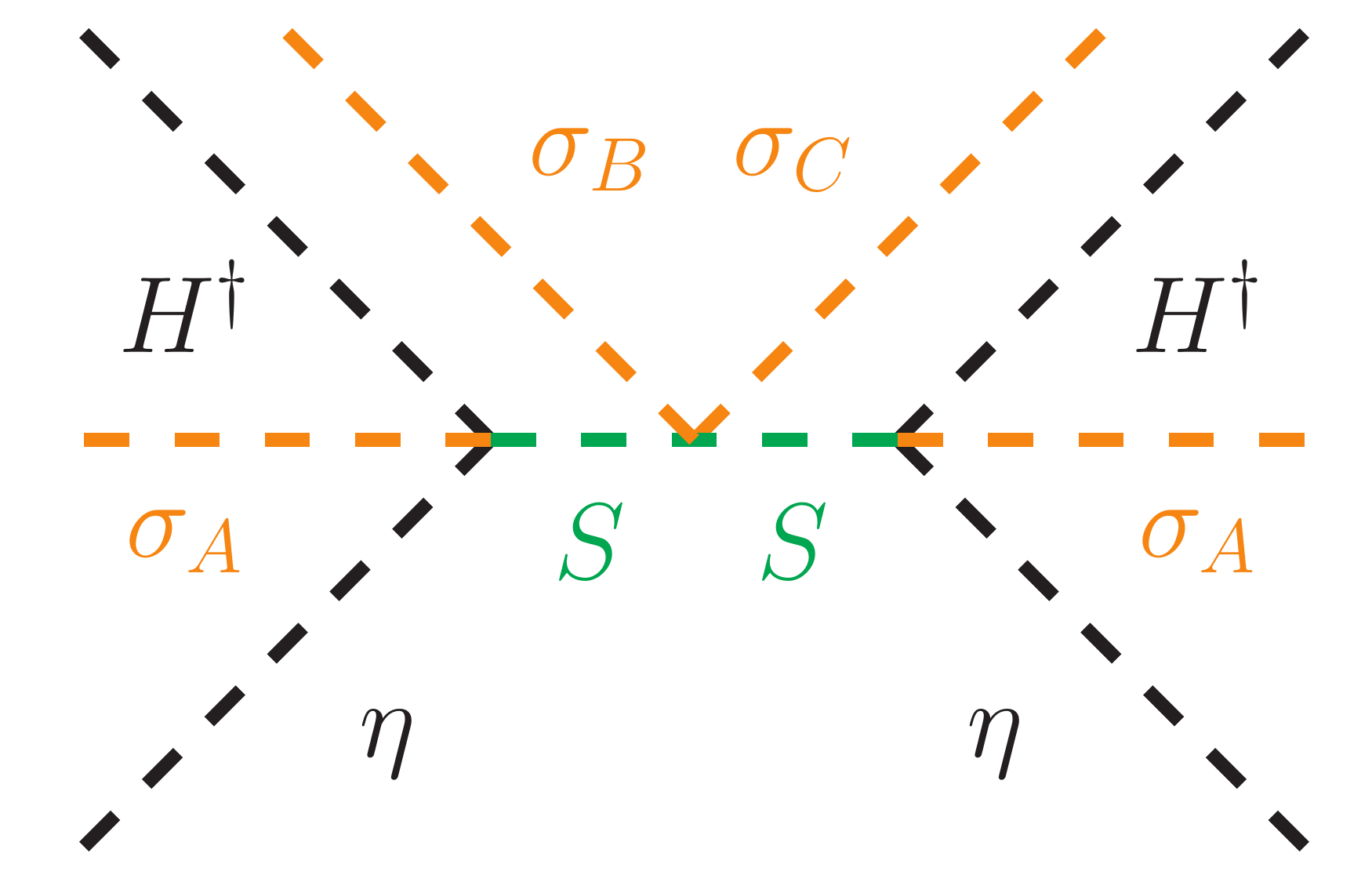}
\end{tabular}&\begin{tabular}{l}
($\sigma_A H^{\dagger}S \eta$),($(S^{\dagger})^2\sigma_B \sigma_C$)
\end{tabular}\\\hline\hline
\end{tabular}
\caption{$\lambda_5$ operator in the UV theory.
  \label{tab:topo}
  }
\end{table}

\noindent We are thus left with neutrino mass topologies with $\sigma$
insertions in both internal lines. The scalar line leads to an
operator $(H^\dagger \eta)^2 \sigma^n$ after the heavy $S$ is
integrated out. All possible topologies are shown in
Table~\ref{tab:topo}. Topologies $\I-\IV$ include one $S$ propagator
and lead to a $\lambda_5$ operator of the form
\begin{equation} \label{eq:lam5IIV}
  \mathcal{O}_{\lambda_5} = (H^\dagger \eta)^2 \sigma_A \sigma_B \, ,
\end{equation}
suppressed by $1/m_S$, while topology $\V$ includes two $S$
propagators and induces the operator
\begin{equation} \label{eq:lam5V}
  \mathcal{O}_{\lambda_5} = (H^\dagger \eta)^2 \sigma_A^2 \sigma_B \sigma_C \, ,
\end{equation}
suppressed by $1/m_S^2$. These two generic expressions for the
$\lambda_5$ operator include cases in which one of the $\sigma$
insertions is missing (for instance, $\sigma_B = \emptyset$) and cases
in which several $\sigma$ insertions in the scalar line correspond to
the same field (for instance, $\sigma_A = \sigma_B$). Finally, the
fermion line simply corresponds to a $\sigma-N-N$ Yukawa
interaction. In the following we will always assume the presence of
the operator $\sigma \overline{N}^c N$ (for models with one $\sigma$
field) or $\sigma_1 \overline{N}^c N$ (for models with two $\sigma$
fields), and we will not draw it. The coefficient of this operator
will be denoted by $\kappa$. Therefore, once the singlet scalar gets a
VEV, $\langle \sigma_{(1)} \rangle =
\frac{v_{\sigma_{(1)}}}{\sqrt{2}}$, the Majorana mass matrix for the
singlet fermions $N$ is generated,~\footnote{In models with two
$\sigma$ fields such that $q_{\sigma_1} = q_{\sigma_2}$ or
$q_{\sigma_1} = - q_{\sigma_2}$, an additional Yukawa term $\sigma_2
\overline{N}^c N$ or $\sigma_2^* \overline{N}^c N$ would be
present. Here $q_{\sigma_1}$ and $q_{\sigma_2}$ denote the $\rm
U(1)_L$ charges of $\sigma_1$ and $\sigma_2$, respectively. This would
lead to $M_N = \sqrt{2} \left( \kappa_1 \, v_{\sigma_1} + \kappa_2 \,
v_{\sigma_2} \right)$ without affecting our discussion. We note,
however, that in such models both $\sigma$ singlets are esentially
copies of the same field.}
\begin{equation}
  M_N = \sqrt{2} \, \kappa \, v_{\sigma_{(1)}} \, .
\end{equation}
In the following we classify all possible UV extensions of the
Scotogenic model compatible with our requirements (A) and (B). Given
their qualitative differences, it is convenient to discuss topologies
$\I-\IV$ and $\V$ separately.

\subsection{Topologies I-IV}
\label{subsec:modelsIIV}

We first discuss the models based on topologies $\I-\IV$. We will
refer to a specific model using the notation $\xi(A,B)$, where $\xi =
\left\{ \I, \II, \III, \IV \right\}$ denotes the topology for the
$(H^\dagger \eta)^2 \sigma_A \sigma_B$ operator, as listed in
Table~\ref{tab:topo}, and $A$ and $B$ denote the singlets involved in
the vertices where $\sigma_{A,B}$ are coupled. Since we only consider
UV theories with at most two different singlets, $A$ and $B$ can only
take the values $\emptyset,1,2,1^*$, where $\emptyset$ indicates that
no $\sigma$ enters the corresponding vertex and $\sigma_{1^*} \equiv
\sigma_1^*$. It is important to mention that we do not consider
scenarios with $A,B=2^*$ because they lead to a redefinition of the
charge $q_{\sigma_2}\to-q_{\sigma_2}$.~\footnote{In the following, we
will denote the $\rm U(1)_L$ charge of the field $X$ as
$q_X$. Furthermore, $q_{\ell_L} = q_{e_R} = 1$ and $q_H = 0$, as
usual.} Therefore, in principle each topology has $16$ different
variations depending on the way the $\sigma_{A,B}$ singlets are
coupled. However, we can reduce this number by taking into account the
following arguments:
\begin{itemize}
\item $A\neq B$ is required to forbid the term $(H^{\dagger}\eta \,
  \sigma_A)^2$ in the effective Lagrangian. If this specific
  combination is allowed, then the term $(H^{\dagger}\eta \,
  \sigma_A)$ is too. This trilinear interaction induces a non-zero VEV
  for $\eta$ after both $H$ and $\sigma_A$ acquire their VEVs, hence
  breaking the Scotogenic \z2 symmetry.
\item $A\neq B^*$ is also required. Otherwise, $(H^{\dagger}\eta)^2
  \sigma_A \sigma_A^*$ is allowed by the $\rm U(1)_L$ symmetry and
  then the operator $(H^{\dagger}\eta)^2$ is present in the UV theory.
\item In all $\xi(1,\emptyset)$ and $\xi(\emptyset,1)$ models the
  effective operator leading to the $\lambda_5$ coupling is
  $\mathcal{O}_{\lambda_5}=(H^{\dagger}\eta)^2\sigma$. This implies
  the relation $2 q_\eta+q_\sigma=0$. In addition, the Yukawa
  coupling $\sigma \overline{N}^c N$ implies $2q_N+q_\sigma=0$. Hence,
  the charges for $\eta$ and $N$ must satisfy $q_\eta=q_N$ and then
  the $\overline{N}\tilde{\eta}^{\dagger}\ell_L$ Yukawa term is
  forbidden by $\rm U(1)_L$. Similarly, in all $\xi(1^*,\emptyset)$
  and $\xi(\emptyset,1^*)$ models one finds $q_\eta=-q_N$ and then
  $q_N=\frac{1}{2}$ in order to allow the term
  $\overline{N}\tilde{\eta}^{\dagger}\ell_L$.
\end{itemize}

\begin{table}[tp]
  \centering
  \renewcommand*{\arraystretch}{1.5}
\begin{tabular}{cccccccccc}\hline\hline
  & {\bf Topology} & $\boldsymbol{A}$ & $\boldsymbol{B}$ & $\boldsymbol{q_N}$ & $\boldsymbol{q_\eta}$ & $\boldsymbol{q_{\sigma_1}}$ & $\boldsymbol{q_{\sigma_2}}$ & $\boldsymbol{q_S}$ & $\boldsymbol{\left(\rm SU(2)_L, \rm U(1)_Y \right)_S}$ \\\hline\hline
  1 & $\I$ & $1^*$ & $\emptyset$ & $\frac{1}{2}$ & $-\frac{1}{2}$ & $-1$ & - & $-1$ & $(\mathbf{3},1)$ \\
  2 & $\I$ & $\emptyset$ & $1^*$ & $\frac{1}{2}$ & $-\frac{1}{2}$ & $-1$ & - & $0$ & $(\mathbf{3},1)$ \\
  3 & $\I$ & $2$ & $\emptyset$ & $q_N$ & $q_N-1$ & $-2 q_N$ & $ 2 - 2q_N$ & $2q_N-2$ & $(\mathbf{3},1)$ \\
  4 & $\I$ & $\emptyset$ & $2$ & $q_N$ & $q_N-1$ & $-2q_N$ & $2-2 q_N$ & $0$ & $(\mathbf{3},1)$ \\
  5 & $\I$ & $1$ & $2$ & $q_N$ & $q_N-1$ & $-2 q_N$ & $2$ & $2 q_N$ & $(\mathbf{3},1)$ \\
  6 & $\I$ & $2$ & $1$ & $q_N$ & $q_N-1$ & $-2 q_N$ & $2$ & $-2$ & $(\mathbf{3},1)$ \\
  7 & $\I$ & $1^*$ & $2$ & $q_N$ & $q_N-1$ & $-2 q_N$ & $2-4q_N$ & $-2 q_N$ & $(\mathbf{3},1)$ \\
  8 & $\I$ & $2$ & $1^*$ & $q_N$ & $q_N-1$ & $-2 q_N$ & $2-4q_N$ & $4q_N-2$ & $(\mathbf{3},1)$ \\
  9-10 & $\II$ & $1^*$ & $\emptyset$ & $\frac{1}{2}$ & $-\frac{1}{2}$ & $-1$ & - & $-\frac{1}{2}$ & $(\mathbf{3},0)$ or $(\mathbf{1},0)$ \\
  11-12 & $\II$ & $2$ & $\emptyset$ & $q_N$ & $q_N-1$ & $-2q_N$ & $2-2q_N$ & $q_N-1$ & $(\mathbf{3},0)$ or $(\mathbf{1},0)$ \\
  13-14 & $\II$ & $1$ & $2$ & $q_N$ & $q_N-1$ & $-2q_N$ & $2$ & $1+q_N$ & $(\mathbf{3},0)$ or $(\mathbf{1},0)$ \\
  15-16 & $\II$ & $1^*$ & $2$ & $q_N$ & $q_N-1$ & $-2q_N$ & $2-4q_N$ & $1-3q_N$ & $(\mathbf{3},0)$ or $(\mathbf{1},0)$ \\
  17 & $\III$ & $1^*$ & $\emptyset$ & $\frac{1}{2}$ & $-\frac{1}{2}$ & $-1$ & - & $-1$ & $(\mathbf{2},1/2)$ \\
  18 & $\III$ & $2$ & $\emptyset$ & $q_N$ & $q_N-1$ & $-2q_N$ & $2-2q_N$ & $2q_N-2$ & $(\mathbf{2},1/2)$ \\
  19 & $\III$ & $1$ & $2$ & $q_N$ & $q_N-1$ & $-2q_N$ & $2$ & $2q_N-2$ & $(\mathbf{2},1/2)$ \\
  20 & $\III$ & $1^*$ & $2$ & $q_N$ & $q_N-1$ & $-2q_N$ & $2-4q_N$ & $2q_N-2$ & $(\mathbf{2},1/2)$ \\
  21 & $\IV$ & $1^*$ & $\emptyset$ & $\frac{1}{2}$ & $-\frac{1}{2}$ & $-1$ & - & $\frac{1}{2}$ & $(\mathbf{2},1/2)$ \\
  22 & $\IV$ & $2$ & $\emptyset$ & $q_N$ & $q_N-1$ & $-2q_N$ & $2- 2q_N$ & $1-q_N$ & $(\mathbf{2},1/2)$ \\
  23 & $\IV$ & $1$ & $2$ & $q_N$ & $q_N-1$ & $-2q_N$ & $2$ & $1-q_N$ & $(\mathbf{2},1/2)$ \\
  24 & $\IV$ & $1^*$ & $2$ & $q_N$ & $q_N-1$ & $-2q_N$ & $2-4q_N$ & $1-q_N$ & $(\mathbf{2},1/2)$ \\
  \hline\hline
\end{tabular}
\caption{
UV extended models leading to topologies $\I-\IV$ and satisfying
conditions (A) and (B). For each model we show the $\rm U(1)_L$
charges of $N$, $\eta$, $\sigma_1$, $\sigma_2$ and $S$, as well as the
$(\rm SU(2)_L, \rm U(1)_Y)$ representation of $S$. Models that become
any of the models in this list after renaming the fields or redefining
their $\rm U(1)_L$ charges are not included, as explained in the text.
\label{tab:clasif}}
\end{table}

\noindent With these considerations, there are only 8 possibilities left in each
of the four topologies. However, there are duplicities. Models based
on topologies $\III$ and $\IV$ are symmetric with respect to the
exchange $\sigma_A\leftrightarrow\sigma_B$ (i.e. $\xi(A,B)=\xi(B,A)$
with $\xi=\III,\IV$). Similarly, $\II(A,B)\sim\II(B,A)$ by redefining
$q_S\to -q_S$. This further reduces the number of fundamentally
different UV models. In total, we find 24 (20 + 4, because in
II-models $S$ can be an $\rm SU(2)_L$ singlet or triplet) different UV
theories. They are listed in Table~\ref{tab:clasif}, where the $\rm
U(1)_L$ charges of $N$, $\eta$, $\sigma_{A,B}$ and $S$, as well as the
$(\rm SU(2)_L, \rm U(1)_Y)$ representation of $S$ in each model, are
shown. Some comments are in order:
\begin{enumerate}[label=(\roman*)]
\item The $(\rm SU(2)_L, \rm U(1)_Y)$ representation of the heavy
  scalar $S$ depends on the topology. In I-models $S$ transforms as
  $(\mathbf{3},1)$, in II-models we have two possibilities,
  $(\mathbf{3},0)$ or $(\mathbf{1},0)$, while in III- and
  IV-models $S$ transforms as $(\mathbf{2},1/2)$.
\item In all the models in Table~\ref{tab:clasif}, the global $\rm
  U(1)_L$ symmetry may be spontaneously broken to a \z2 parity, under
  which $N$ and $\eta$ are odd. In all the $\xi(1^*,\emptyset)$ models
  and in $\I(\emptyset,1^*)$, the conservation of $\rm U(1)_L$
  restricts the lepton number charges of $N$, $\eta$, $\sigma_{A,B}$
  and $S$, which must take precise values, and this automatically
  implies a remnant \z2 that corresponds to the usual Scotogenic
  parity. The model studied in Ref.~\cite{Escribano:2021ymx}, which
  corresponds to model $\I(1^*,\emptyset)$ in our notation, is a good
  example of this. In the rest of the models, the conservation of $\rm
  U(1)_L$ leaves one of the charges to be chosen freely. We decided to
  choose $q_N$. In this case, these are the restrictions to recover
  the \textit{dark} \z2 parity from $\rm U(1)_L$ breaking:
\begin{itemize}
\item $q_N$ cannot be an integer. 
\item If $q_N=\frac{\alpha}{\beta}$, with $\alpha,\beta \in \mathbb{Z}$, then $\alpha$ and $\beta$ have to be odd and even, respectively.
\item $\GCD(\alpha,\beta)=1$, where GCD stands for Greatest Common Divisor. Therefore, $\alpha$ and $\beta$ must be coprime.
\end{itemize}
The first restriction comes from the requirement of $N$ and $\eta$
being both odd under the remnant Scotogenic \z2. The relation
$q_\eta=q_N-1$ implies that if $q_N$ is even, then $q_\eta$ must be
odd, and vice versa. Then, $N$ and $\eta$ will transform differently
under the remnant \z2 symmetry. As an example of this consider the
model $\I(1,2)$ with $q_N=2$. In this case, the solution for the rest
of the $\rm U(1)_L$ charges in the model is $q_\eta=1,
q_{\sigma_1}=-4, q_{\sigma_2}=2$ and $q_S=4$. The global lepton number
symmetry gets spontaneously broken as $\rm U(1)_L\to \mathbb{Z}_2$,
but with $N$ and $\eta$ charged under \z2 as $+$ and $-$,
respectively, and this does not reproduce the Scotogenic
model. Similarly, if $q_N=\frac{\alpha}{\beta}$, after normalizing all
$\rm U(1)_L$ charges so that they become integer numbers (multiplying
by $\beta$) we obtain $\tilde{q}_\eta=\beta-\alpha$ and
$\tilde{q}_N=\alpha$. Hence, for $\eta$ and $N$ to be odd under \z2,
$\alpha$ and $\beta$ must be odd and even, respectively. Finally, the
third restriction is required to guarantee that $n=2$ after $\rm
U(1)_L$ breaks to the discrete symmetry $\mathbb{Z}_n$. As an example
we take model I(1,2), where $n \equiv
\GCD(\tilde{q}_{\sigma_1},\tilde{q}_{\sigma_2},\tilde{q}_S) =
\GCD(-2\alpha,2\beta,2\alpha) = 2 \GCD(\alpha,\beta)= 2$. We checked
for all the working models that
$\GCD(\tilde{q}_{\sigma_1},\tilde{q}_{\sigma_2},\tilde{q}_S)$ or
$\GCD(\tilde{q}_{\sigma_1},\tilde{q}_{\sigma_2})$, depending on
whether $S$ acquires a VEV or not, always reduces to $\GCD(\alpha,
\beta) = 1$. Also, we want $q_N = \frac{\alpha}{\beta}$ to be
irreducible.
\item In all models, and for all possible values of $q_N$ in agreement
  with the restrictions listed in the previous item, $\eta$ never
  acquires an induced VEV. This is crucial for the consistency of the
  Scotogenic model.
\item It is clear that in all models of the form $\xi(A,\emptyset)$ or
  $\xi(\emptyset,B)$, a trilinear coupling $\mu$ participates in the
  generation of the $\lambda_5$ coupling, induced after the breaking
  of $\rm U(1)_L$. This is perfectly consistent, but requires the
  assumption $\mu \ll m_S$ to justify $\lambda_5 \ll 1$. This poses a
  theoretical issue, since $\mu$ is a parameter of the UV theory. In
  contrast, in models of the form $\xi(A,B)$ with $A,B\neq \emptyset$,
  the $\lambda_5$ coupling will only depend on the $\sigma_{A,B}$
  VEVs, induced at low energies and naturally small compared to $m_S$.
\item Finally, we note that in I-models the $\rm U(1)_L$ charges of
  the particles $N$, $\eta$ and $\sigma_{A,B}$ remain the same after
  the non-trivial change $A \leftrightarrow B$. For instance, this is
  the case in models $\I(1,2)$ and $\I(2,1)$.
\end{enumerate}

\subsection{Topology V}
\label{subsecs:modelsV}

Topology $\V$ contains an internal line with a double $S$ propagator
and thus induces the $\lambda_5$ coupling at order $1/m_S^4$. This is
two orders higher than the corresponding contributions from topologies
$\I-\IV$. Therefore, for a diagram with topology $\V$ to be dominant,
other topologies must be absent (or highly suppressed due to a
specific parameter choice). In fact, many models leading to topology
$\V$ also generate other topologies, and they have been already
included in the previous discussion. Nevertheless, there are also some
models in which the symmetries allow for topology $\V$ but forbid
other topologies, as we proceed to show.

Topology $\V$ requires the presence of the operators $H^\dagger \eta
\, S \, \sigma_A$ and $\left(S^\dagger\right)^2 \sigma_B \sigma_C$ to
produce a $\lambda_5$ operator as in Eq.~\eqref{eq:lam5V}. A model
based on this topology will be denoted as $\V(A,B,C)$, where $A$, $B$,
and $C$ can take the values $\emptyset,1,2,1^*$, as in the previous
topologies. Again, we do not consider models with $A, B, C = 2^*$. The
reason, however, is twofold. On the one hand, in scenarios involving
$2^*$ but not $2$ this is again due to the existence of a redefinition
of the charges that allows to show an equivalence to models without
$2^*$. On the other hand, models combining $2$ and $2^*$ do not lead to a solution for the $U(1)_{\rm L}$ charges or their solutions are compatible with topology $\II$, which is naturally dominant.~\footnote{This is the
case of models $\V(2,2^*,C)$ and $\V(2,B,2^*)$. These are not
equivalent to models $\V(2,2,C)$ and $\V(2,B,2)$, respectively, so
they do not lead to just a redefinition of the charges.} In
conclusion, topology $\V$ leads to $4\times 4\times 4 = 64$ different
variations depending on the way the $\sigma_{A,B,C}$ singlets are
coupled. However, we can reduce this number by taking into account the
following arguments:
\begin{itemize}
\item All $\V$ models are symmetric under $B \leftrightarrow C$,
  $\V(A,B,C) = \V(A,C,B)$. Then, for each of the 4 possible values of
  $A$, this removes $(4^2 - 4)/2 = 6$ possibilities, leaving $40$
  variations.
\item $B\neq C$ and $B\neq C^*$. The former is required to forbid the
  operator $S^{\dagger} \sigma_B$. This would induce a VEV for $S$,
  which in turn would induce a VEV for $\eta$ due to the operator
  $H^{\dagger} \eta \, S \, \sigma_A$. The latter restriction avoids
  having $S^{\dagger} S^{\dagger}$ in the Lagrangian, since this term
  would imply that $S$ is a singlet under every symmetry of the model,
  hence leading to an induced VEV for $\eta$ as well. This condition
  together with the previous one leaves $4 \times \left[ (4^2 + 4)/2 -
    4 - 1 \right] = 20$ possibilities.
\item $A \neq B^*$ (or $C^*$) leads either to models for which the
  equations for the charges are incompatible with the original
  assumptions or to models for which the solutions for the charges are
  compatible with topology $\II$. Here $\mathcal{O}_{\lambda_5}$ takes
  the form $(\sigma_A^* \sigma_A)\sigma_A
  \sigma_C(H^{\dagger}\eta)^2$, which means that the term $\sigma_A
  \sigma_C(H^{\dagger}\eta)^2$ would be allowed by lepton number and,
  in turn, $\sigma_C (H^{\dagger}S^{\dagger}\eta)$ too, with $C\neq A$
  in order to satisfy the above requirements. Given that, by
  construction, we have the operator $\sigma_A(H^{\dagger}S\eta)$
  within the model, the diagram for the scalar line of II-models
  (shown in Table~\ref{tab:topo}) would appear, leaving this new
  topology as a subdominant effect in the generation of the
  $\lambda_5$ coupling. From the remaining 20 variations, this removes
  2 for each $A = 1$ and $A = 1^*$, and 3 more for the models
  $\V(\emptyset,B,\emptyset)$, leaving 13 possibilities.
\end{itemize}
Notice that $S$ can be a singlet and a triplet in all the models, so
we have the $26$ models shown in Table~\ref{tab:clasifV}. Again, we
note that if only one of the three A, B, or C labels is equal to $2$,
then the same model but with $2^*$ instead, is equivalent to the
former with $q_{\sigma_2} \rightarrow - q_{\sigma_2}$.

\begin{table}[tp]
  \centering
  \renewcommand*{\arraystretch}{1.5}
\begin{tabular}{ccccccccccc}\hline\hline
  & {\bf Topology} & $\boldsymbol{A}$ & $\boldsymbol{B}$ & $\boldsymbol{C}$ & $\boldsymbol{q_N}$ & $\boldsymbol{q_\eta}$ & $\boldsymbol{q_{\sigma_1}}$ & $\boldsymbol{q_{\sigma_2}}$ & $\boldsymbol{q_S}$ & $\boldsymbol{\left(\rm SU(2)_L, \rm U(1)_Y \right)_S}$ \\\hline\hline
  25-26 & $\V$ & $1$ & $1$ & $\emptyset$ & $-\frac{1}{2}$ & $-\frac{3}{2}$ & $1$ & - & $\frac{1}{2}$ & $(\mathbf{3},0)$ or $(\mathbf{1},0)$ \\
  27-28 & $\V$ & $1^*$ & $1^*$ & $\emptyset$ & $\frac{1}{4}$ & $-\frac{3}{4}$ & $-\frac{1}{2}$ & - & $\frac{1}{4}$ & $(\mathbf{3},0)$ or $(\mathbf{1},0)$ \\
  29-30 & $\V$ & $\emptyset$ & $1$ & $2$ & $q_N$ & $q_N-1$ & $-2q_N$ & $2$ & $1-q_N$ & $(\mathbf{3},0)$ or $(\mathbf{1},0)$ \\
  31-32 & $\V$ & $\emptyset$ & $1^*$ & $2$ & $q_N$ & $q_N-1$ & $-2q_N$ & $2-4q_N$ & $1-q_N$ & $(\mathbf{3},0)$ or $(\mathbf{1},0)$ \\
  33-34 & $\V$ & $1$ & $2$ & $\emptyset$ & $q_N$ & $q_N-1$ & $-2q_N$ & $2q_N+2$ & $1+q_N$ & $(\mathbf{3},0)$ or $(\mathbf{1},0)$ \\
  35-36 & $\V$ & $1^*$ & $2$ & $\emptyset$ & $q_N$ & $q_N-1$ & $-2q_N$ & $2-6q_N$ & $1-3q_N$ & $(\mathbf{3},0)$ or $(\mathbf{1},0)$ \\
  37-38 & $\V$ & $1$ & $1$ & $2$ & $q_N$ & $q_N-1$ & $-2q_N$ & $4q_N+2$ & $1+q_N$ & $(\mathbf{3},0)$ or $(\mathbf{1},0)$ \\
  39-40 & $\V$ & $1^*$ & $1^*$ & $2$ & $q_N$ & $q_N-1$ & $-2q_N$ & $2-8q_N$ & $1-3q_N$ & $(\mathbf{3},0)$ or $(\mathbf{1},0)$ \\
  41-42 & $\V$ & $2$ & $1$ & $\emptyset$ & $q_N$ & $q_N-1$ & $-2q_N$ & $1$ & $-q_N$ & $(\mathbf{3},0)$ or $(\mathbf{1},0)$ \\
  43-44 & $\V$ & $2$ & $1^*$ & $\emptyset$ & $q_N$ & $q_N-1$ & $-2q_N$ & $1-2q_N$ & $q_N$ & $(\mathbf{3},0)$ or $(\mathbf{1},0)$ \\
  45-46 & $\V$ & $2$ & $2$ & $\emptyset$ & $q_N$ & $q_N-1$ & $-2q_N$ & $\frac{2}{3}-\frac{2}{3}q_N$ & $\frac{1}{3}-\frac{1}{3}q_N$ & $(\mathbf{3},0)$ or $(\mathbf{1},0)$ \\
  47-48 & $\V$ & $2$ & $1$ & $2$ & $q_N$ & $q_N-1$ & $-2q_N$ & $\frac{2}{3}$ & $\frac{1}{3}-q_N$ & $(\mathbf{3},0)$ or $(\mathbf{1},0)$ \\
  49-50 & $\V$ & $2$ & $1^*$ & $2$ & $q_N$ & $q_N-1$ & $-2q_N$ & $\frac{2}{3}-\frac{4}{3}q_N$ & $\frac{1}{3}+\frac{1}{3}q_N$ & $(\mathbf{3},0)$ or $(\mathbf{1},0)$ \\
  \hline\hline
\end{tabular}
\caption{  
UV extended models leading to topology $\V$ and satisfying conditions
(A) and (B). For each model we show the $\rm U(1)_L$ charges of $N$,
$\eta$, $\sigma_1$, $\sigma_2$ and $S$, as well as the $(\rm SU(2)_L,
\rm U(1)_Y)$ representation of $S$. Models that become any of the
models in this list after renaming the fields or redefining their $\rm
U(1)_L$ charges are not included, as explained in the text.
\label{tab:clasifV}}
\end{table}

Again, in all the models in Table~\ref{tab:clasifV}, the global $\rm
U(1)_L$ symmetry may be spontaneously broken to a \z2 parity, under
which $N$ and $\eta$ are odd. In the models $\V(1,1,\emptyset)$ and
$\V(1^*,1^*,\emptyset)$, the conservation of $\rm U(1)_L$ restricts
the lepton number charges of $N$, $\eta$, $\sigma_{A,B,C}$ and $S$,
which must take precise values, and this automatically implies a
remnant \z2 that corresponds to the usual Scotogenic parity. In the
rest of the models, the conservation of $\rm U(1)_L$ leaves one of the
charges to be chosen freely. We decided to choose $q_N$. In this case,
these are the restrictions to recover the \textit{dark} \z2 parity
from $\rm U(1)_L$ breaking:
\begin{itemize}
\item $q_N$ cannot be an integer. 
\item If $q_N=\frac{\alpha}{\beta}$, with $\alpha,\beta \in \mathbb{Z}$, then $\alpha$ and $\beta$ have to be odd and even, respectively.
\item $\GCD(\alpha,\beta)=1$. Therefore, $\alpha$ and $\beta$ must be coprime.
\end{itemize}
Additionally, some models have extra conditions for the \z2 to appear:
\begin{itemize}
\item In $\V(2,2,0)$, we further require $\GCD(3 \, \alpha,\alpha - \beta)= 1$ if $\frac{\alpha - \beta}{3}$ is not an integer, or $\GCD(\alpha,\frac{\alpha - \beta}{3})= 1$ if $\frac{\alpha - \beta}{3}$ is an integer.
\item In $\V(2,1^*,2)$, we further require $\GCD(3 \, \alpha,2\,\alpha - \beta)= 1$ if $\frac{2 \, \alpha - \beta}{3}$ is not an integer, or $\GCD(\alpha,\frac{2\, \alpha - \beta}{3})= 1$ if $\frac{2\, \alpha - \beta}{3}$ is an integer.
\item In $\V(2,1,2)$, we further require $\GCD(3 \, \alpha,\beta)= 1$ if $\frac{\beta}{3}$ is not an integer, or $\GCD(\alpha, \frac{\beta}{3}) = 1$ if $\frac{\beta}{3}$ is an integer.
\end{itemize}
This concludes our classification of all possible UV extensions of the
Scotogenic model satisfying our requirements (A) and (B). We will now
illustrate it with two specific example models. An additional example
can be found in~\cite{Escribano:2021ymx}.

\section{An UV extended Scotogenic model with one $\sigma$ field}
\label{sec:model1}

Our first example model is an UV extension of the Scotogenic model
with one $\sigma$ field. Another example of this class of models can
be found in~\cite{Escribano:2021ymx}.

\subsection{Ultraviolet theory}

We consider an extension of the Scotogenic model with two new
particles: the $\rm SU(2)_L$ doublet $S$ and the singlet $\sigma$,
both scalars. The \z2 Scotogenic parity is replaced by a global $\rm
U(1)_L$ lepton number symmetry. Table~\ref{tab:ParticleContent1} shows
the scalar and leptonic fields of the model and their representations
under the gauge and global symmetries.

{
\renewcommand{\arraystretch}{1.4}
\begin{table}[tb]
  \centering
  \begin{tabular}{|c|c||ccc|c|}
      \hline
      \textbf{Field} & \textbf{Generations} & \textbf{$\rm SU(3)_c$} & \textbf{$\rm SU(2)_L$} & \textbf{$\rm U(1)_Y$} & \textbf{$\rm U(1)_L$} \\
      \hline
      $\ell_L$ &  3  &  \one  &  \two  &  -1/2  &  1   \\
      $e_R$    &  3  &  \one  &  \one  &  -1    &  1   \\
      $N$      &  3  &  \one  &  \one  &  0     &  $q_N$  \\
      \hline
      $H$      &  1  &  \one  &  \two  &  1/2   &  0   \\
      $\eta$   &  1  &  \one  &  \two  &  1/2   &  $q_\eta$  \\
      $\sigma$ &  1  &  \one  &  \one  &  0     &  $q_\sigma$  \\
      $S$      &  1  &  \one  &  \two  &  1/2   &  $q_S$   \\
      \hline
  \end{tabular}
  \caption{\label{tab:ParticleContent1}
    Lepton and scalar particle content and representations under the gauge and global symmetries in an UV extension of the Scotogenic model with one $\sigma$ field.}
\end{table}
}

\begin{figure}[tb!]
  \centering
  \includegraphics[width=0.5\linewidth]{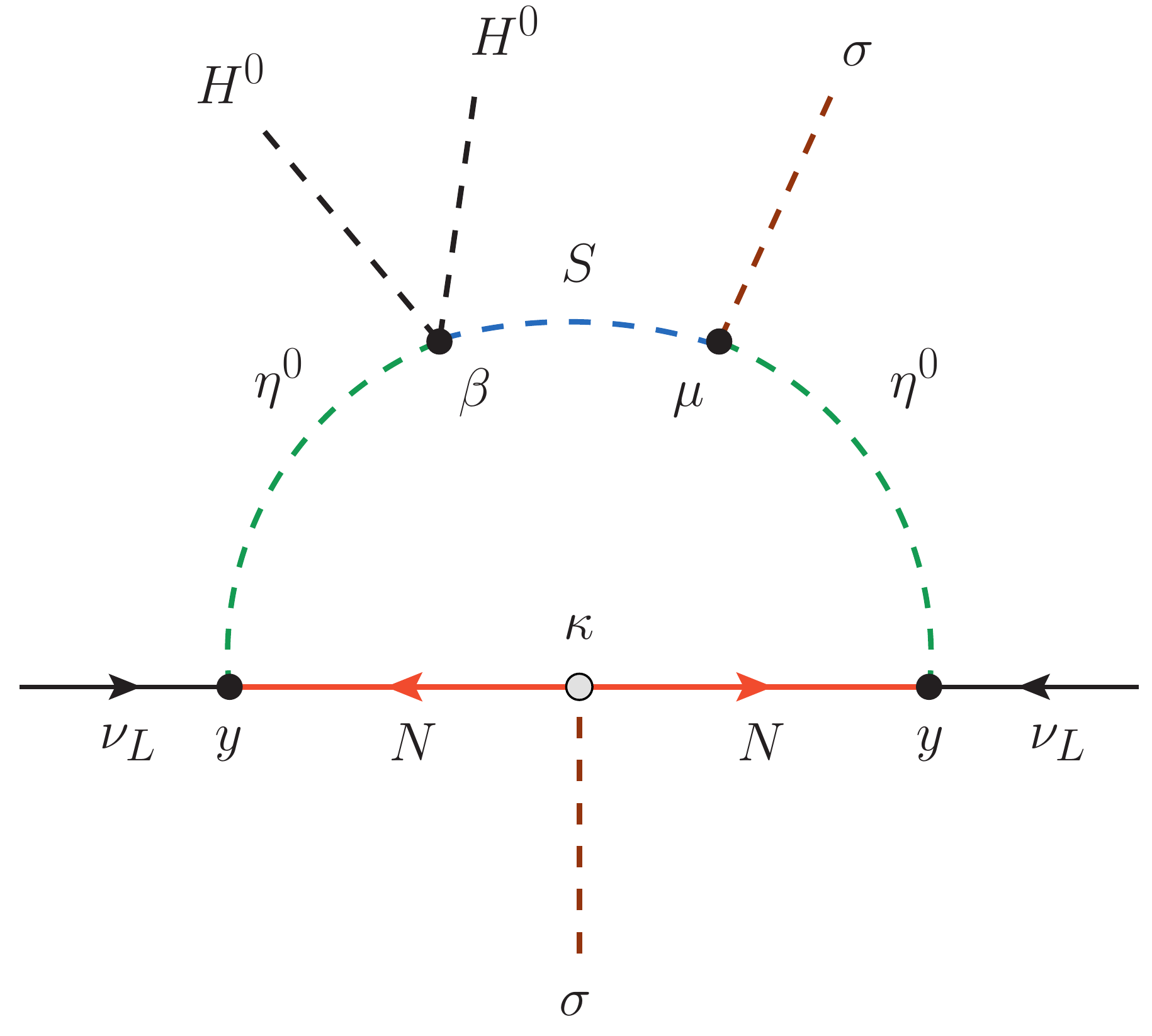}
  \caption{Neutrino mass generation in an extended Scotogenic model
    with one $\sigma$ field. This Feynman diagram shows the relevant
    gauge eigenstates involved in the 1-loop contribution to neutrino
    masses. In our notation, this is a $\IV(1^*,\emptyset)$ model.
    \label{fig:scot1}
    }
\end{figure}

We want to explain the smallness of the Scotogenic's $\lambda_5$
coupling. Our strategy will be to forbid it in our original Lagrangian
and make it arise effectively at low energies once the scalar $\sigma$
acquires a VEV and we integrate out $S$. We also impose that, after
symmetry breaking, the effective $\lambda_5$ coupling would induce
neutrino masses as shown in Fig.~\ref{fig:scot1}. In our notation,
this is a $\IV(1^*,\emptyset)$ model. This requires the presence of
the operators
\begin{equation}
  \overline{N} \widetilde{\eta}^\dagger \ell_L \quad , \quad \sigma \overline{N}^c N \quad , \quad H^\dagger S H^\dagger \eta \quad , \quad \sigma^* S^\dagger \eta \, ,
\end{equation}
which in turn imply the following set of equations for the $\rm
U(1)_L$ charges of the model:
\begin{align}
  -q_N + q_\eta + 1 &= 0 \, , \label{eq:sys1} \\
  q_\sigma + 2 \, q_N &= 0 \, , \label{eq:sys2} \\
  q_S + q_\eta &= 0 \, , \label{eq:sys3} \\
  -q_\sigma - q_S + q_\eta &= 0 \, . \label{eq:sys4}
\end{align}
This system of linear equations has a unique solution:
\begin{align}
  q_N &= \frac{1}{2} \, , \\
  q_\eta &= - \frac{1}{2} \, , \\
  q_\sigma &= -1 \, , \\
  q_S &= \frac{1}{2} \, .
\end{align}
With this solution, the operators
\begin{equation}
  \overline{N}^c N \quad , \quad
  \overline{N} \widetilde{H}^\dagger \ell_L \quad , \quad
  \left( H^\dagger \eta \right)^2
\end{equation}
are automatically forbidden due to $\rm U(1)_L$ conservation. One
should note that if we chose the operator $\sigma S^\dagger \eta$
instead of $\sigma^\ast S^\dagger \eta$, no solution for the resulting
system of equations would exist. Indeed, if one replaces $-q_\sigma$
by $q_\sigma$ in Eq.~\eqref{eq:sys4}, the combination of the resulting
equation with Eqs.~\eqref{eq:sys2} and \eqref{eq:sys3} leads to $q_N =
q_\eta$, which is incompatible with Eq.~\eqref{eq:sys1}. This
illustrates why $\xi(1,\emptyset)$ models are not compatible with our
requirements.

Having fixed the quantum numbers of all the particles in the model, we
proceed to write its Lagrangian. The new Yukawa interactions are given
by
\begin{equation}
  \mathcal{L}_{\rm Y} = y \, \overline{N} \, \widetilde{\eta}^\dagger \, \ell_L + \kappa \, \sigma \overline{N}^c N + \hc \, ,
\end{equation}
where $y$ is a general complex $3 \times 3$ matrix and $\kappa$ is a
complex symmetric $3 \times 3$ matrix. The scalar potential of the
model can be written as
\begin{equation}
\begin{split}
\mathcal{V}_{\rm UV} &= m_H^2 H^{\dagger} H+m_S^2 S^{\dagger} S+m_\sigma^2 \sigma^* \sigma+ m_{\eta}^2 \eta^{\dagger}\eta +\frac{\lambda_1}{2}(H^{\dagger}H)^2+\frac{\lambda_2}{2} (\eta^{\dagger} \eta)^2 \\
&+\frac{\lambda_S}{2}(S^{\dagger} S)^2+\frac{\lambda_\sigma}{2}(\sigma^*\sigma)^2 + \lambda_3 (H^{\dagger}H)(\eta^{\dagger}\eta)+\lambda_3^S(H^{\dagger}H)(S^{\dagger} S) \\
&+\lambda_3^\sigma(H^{\dagger}H)(\sigma^{\dagger} \sigma)+\lambda_3^{\eta S}(\eta^\dagger \eta)(S^{\dagger} S)+\lambda_3^{\eta \sigma}(\eta^\dagger \eta)(\sigma^*\sigma) \\
&+\lambda_3^{\sigma S}(\sigma^*\sigma)(S^{\dagger} S)+\lambda_4 (H^\dagger \eta)(\eta^\dagger H)+\lambda_4^{HS} (H^{\dagger}S)(S^{\dagger}H)\\
&+\lambda_4^{\eta S}(S^{\dagger}\eta)(\eta^{\dagger}S) +\left[\beta(H^{\dagger} S H^{\dagger} \eta)+ \mu (\sigma^* S^{\dagger} \eta)+ \hc \right] \, .
\end{split}\label{UUV1}
\end{equation}
Here $\mu$ is a trilinear parameter with dimensions of mass while
$m_H^2$, $m_\eta^2$ and $m_\sigma^2$ have dimensions of
mass$^2$. Other Lagrangian terms are allowed by the gauge symmetries
of the model but forbidden by $\rm U(1)_L$.

\subsection{Effective theory}

We will now assume that $m_S$ is much larger than any other energy
scale in the theory. At energies well below $m_S$, all physical
processes can be properly described by an effective field theory in
which the heavy field $S$ has been integrated out. We now present this
effective theory, obtained after integrating out $S$ at
tree-level. The effective potential at low energies can be written as
\begin{equation}
\begin{split}
\mathcal{V}_{\rm IR}&=m_H^2 H^{\dagger}H+m_{\eta}^2\eta^{\dagger}\eta+m_{\sigma}^2\sigma^*\sigma+\frac{\lambda_1}{2}(H^{\dagger}H)^2+\frac{\lambda_2}{2}(\eta^{\dagger}\eta)^2+\frac{\lambda_{\sigma}}{2}(\sigma^*\sigma)^2\\
&+\lambda_3 (H^{\dagger}H)(\eta^{\dagger}\eta)+\lambda_3^{\sigma}(H^{\dagger}H)(\sigma^*\sigma)+\left[\lambda_3^{\eta\sigma}-\frac{|\mu|^2}{m_S^2}\right](\sigma^*\sigma)(\eta^{\dagger}\eta)\\
&+\left[\lambda_4-\frac{|\beta|^2(H^{\dagger}H)}{m_S^2}\right](H^{\dagger}\eta)(\eta^{\dagger}H)-\left[\frac{\beta \mu}{m_S^2} \sigma^* (H^{\dagger}\eta)^2 + \hc \right] +\mathcal{O}\left(\frac{1}{m_S^4}\right) \, .
\end{split}\label{VIRmodel1}
\end{equation}
Assuming that CP is conserved in the scalar sector, the neutral fields
$H^0$ and $\sigma$ can be decomposed as
\begin{equation}
H^0=\frac{1}{\sqrt{2}}(v_H+\phi+i A) \, , \quad \sigma=\frac{1}{\sqrt{2}}(v_{\sigma}+\rho+i J) \, ,
\end{equation}
with $\frac{v_H}{\sqrt{2}}$ and $\frac{v_{\sigma}}{\sqrt{2}}$ the VEVs
of $H^0$ and $\sigma$, respectively. These VEVs are determined by
minimizing the scalar potential in Eq.~\eqref{VIRmodel1}. The
resulting tadpole equations are given by
\begin{align}
\left. \frac{d \mathcal{V}_{\rm IR}}{d H^0}\right|_{\langle H^0, \sigma\rangle=\{\frac{v_H}{\sqrt{2}},\frac{v_{\sigma}}{\sqrt{2}}\}}=&\frac{v_H}{\sqrt{2}}\left( m_{H}^2+\frac{\lambda_1 v_H^2}{2}+\frac{\lambda_3^{\sigma}v_{\sigma}^2}{2}\right) \, , \label{tp1m1}\\
\left. \frac{d \mathcal{V}_{\rm IR}}{d \sigma}\right|_{\langle H^0,\sigma\rangle=\{\frac{v_H}{\sqrt{2}},\frac{v_{\sigma}}{\sqrt{2}}\}}=&\frac{v_{\sigma}}{\sqrt{2}}\left( m_{\sigma^2}+\frac{\lambda_3^{\sigma} v_H^2}{2}+\frac{\lambda_{\sigma}v_{\sigma}^2}{2}\right) \, , \label{tp2m1}
\end{align}
where we have only written the non-trivial equations and these are
evaluated at the VEVs of each scalar field.  As we see from
Eq.~\eqref{VIRmodel1}, once $\sigma$ acquires a VEV, the operator
$(H^{\dagger}\eta)^2$ is generated, with an effective $\lambda_5$
coupling that is naturally suppressed by the mass of the heavy field
$S$,
\begin{equation} \label{eq:lambda51}
\frac{\lambda_5}{2}=-\frac{\beta \mu v_{\sigma}}{\sqrt{2}m_S^2} \ll 1 \, .
\end{equation}
This follows from the assumption $\mu \ll m_S$. As explained in
Sec.~\ref{sec:clasif}, this is perfectly valid. However, it poses a
theoretical problem since $\mu$ is parameter of the UV theory. A model
without this issue will be discussed below in
Sec.~\ref{sec:model2}. We now proceed to the computation of the scalar
spectrum of the model. In the bases $\{\phi,\rho\}$ for the CP-even
states and $\{A,J\}$ for the CP-odd ones, the squared mass matrices
read
\begin{equation}
\mathcal{M}_{R}^2=\left(\begin{array}{cc}
m_{H}^2+\frac{1}{2}\left(3\lambda_1 v_H^2+\lambda_3^{\sigma}v_{\sigma}^2 \right)& \lambda_3^{\sigma}v_H v_{\sigma}\\
\lambda_3^{\sigma}v_H v_{\sigma} & m_{\sigma}^2+\frac{1}{2}\left(\lambda_3^{\sigma}v_{H}^2+3\lambda_{\sigma}v_{\sigma}^2\right)
\end{array}\right) \, ,
\end{equation}
and
\begin{equation}
\mathcal{M}_{I}^2=\left(\begin{array}{cc}
m_H^2+\frac{\lambda_1 v_{H}^2}{2}+\frac{\lambda_3^{\sigma}v_{\sigma}^2}{2} & 0\\
0 & m_{\sigma}^2+\frac{\lambda_{\sigma}^2 v_H^2}{2}+\frac{\lambda_{\sigma} v_{\sigma}^2}{2}
\end{array}\right) \, ,
\end{equation}
respectively. The above expressions can be reduced using
Eqs.~\eqref{tp1m1} and \eqref{tp2m1}. When this is done, the resulting
$\mathcal{M}_{I}^2$ becomes identically zero. This implies the
existence of two massless Goldstone bosons. One of them ($A$)
corresponds to the state that is \textit{eaten up} by the $Z$ boson
and becomes its longitudinal component, while the other ($J$) is
associated to the spontaneous breaking of the global $\rm U(1)_L$
symmetry, the so-called majoron. On the other hand, the reduction of
$\mathcal{M}_{R}^2$ with Eqs.~\eqref{tp1m1} and \eqref{tp2m1} leads to
\begin{equation}
\mathcal{M}_{R}^2=\left(\begin{array}{cc}
\lambda_1 v_{H}^2& \lambda_3^{\sigma}v_H v_{\sigma}\\
\lambda_3^{\sigma}v_H v_{\sigma} & \lambda_{\sigma}v_{\sigma}^2
\end{array}\right) \, .
\end{equation}
This matrix can be brought to diagonal form as $ V_R^T \mathcal{M}_R^2
V_R = \widehat{\mathcal{M}}_R^2 = \text{diag}(m_h^2,m_{\Phi}^2)$,
where $V_R$ is a unitary matrix that can be parametrized as
\begin{equation}
V_R=\left(\begin{array}{cc}
\cos \theta &-\sin \theta\\
\sin \theta & \cos \theta
\end{array}\right) \, .
\end{equation}
The mixing angle $\theta$ is given by
\begin{equation}
\tan(2\theta)=\frac{2(\mathcal{M}_R^2)_{12}}{(\mathcal{M}_R^2)_{11}-(\mathcal{M}_R^2)_{22}}=\frac{2 r\lambda_{3}^{\sigma}}{r^2 \lambda_1-\lambda_{\sigma}}\approx -2r\frac{\lambda_3^{\sigma}}{\lambda_{\sigma}}+\mathcal{O}(r^2) \, ,
\end{equation}
with $r\equiv v_H/v_{\sigma}$. For $v_\sigma \sim$ TeV, $r \ll 1$ and
simple approximate expressions can be obtained. The lightest of the
two mass eigenstates is the well-known Higgs-like state $h$, with mass
$m_h\approx125$ GeV, discovered at the LHC. In addition, the model
contains the heavy scalar $\Phi$, with a mass of the order of
$v_\sigma$. We focus now on the \z2-odd scalars $\eta^+$ and
$\eta^0$. The neutral $\eta^0$ field can be decomposed as
\begin{equation}
\eta^0=\frac{1}{\sqrt{2}}(\eta_R+i \eta_I) \, .
\end{equation}
Their masses are given by
\begin{align}
m_{\eta^+}&=m_{\eta}^2+\frac{v_H^2}{2}\lambda_3^{\rm eff} \, ,\\
m_{\eta_R}^2&=m_{\eta}^2+\frac{v_H^2}{2}\left(\lambda_3^{\rm eff} +\lambda_4^{\rm eff}-\sqrt{2} \, \frac{\beta \mu v_{\sigma}}{m_S^2}\right) \, ,\\
m_{\eta_I}^2&=m_{\eta}^2+\frac{v_H^2}{2}\left(\lambda_3^{\rm eff} +\lambda_4^{\rm eff}+\sqrt{2} \, \frac{\beta \mu v_{\sigma}}{m_S^2}\right) \, ,
\end{align}
where we have defined
\begin{align}
\lambda_3^{\rm eff}&\equiv \lambda_3+\lambda_3^{\eta\sigma}\frac{v_{\sigma}^2}{v_H^2}-\mu^2 \frac{v_{\sigma}^2}{v_H^2 m_S^2} \, ,\\
\lambda_4^{\rm eff}&\equiv \lambda_4-\frac{\beta^2 v_H^2}{2 m_S^2} \, .
\end{align}
The mass square difference between $\eta_R$ and $\eta_I$ is given by
\begin{equation}
m_{\eta_R}^2-m_{\eta_I}^2=-\sqrt{2} \, \frac{\beta\mu v_{\sigma}}{m_S^2}v_H^2=\lambda_5 v_H^2 \, ,
\end{equation}
exactly as in the usual Scotogenic model. Finally, the spontaneous
breaking of $\rm U(1)_L$ by the VEV of $\sigma$ induces a Majorana
mass term for the $N$ singlets, with $M_N = \sqrt{2} \, \kappa \,
v_\sigma$. This leads to Majorana neutrino masses at 1-loop, as shown
in Fig.~\ref{fig:scot1}. The $3 \times 3$ neutrino mass matrix is
given by usual Scotogenic formula in Eq.~\eqref{eq:numass}, where
$\lambda_5$ is the effective coupling in Eq.~\eqref{eq:lambda51}. Due
to the additional scalar states, including a massless majoron with
couplings to charged leptons, the phenomenology of this model is
richer than that of the usual Scotogenic scenario. This will be
discussed in Sec.~\ref{sec:pheno}.

\section{An UV extended Scotogenic model with two $\sigma$ fields}
\label{sec:model2}

We consider now an UV extension of the Scotogenic model with two $\sigma$
fields.

\subsection{Ultraviolet theory}

We enlarge the Scotogenic particle content with three new particles:
the scalar $\rm SU(2)_L$ singlets $S$, $\sigma_1$ and
$\sigma_2$. Again, instead of the usual \z2 Scotogenic parity, a
global $\rm U(1)_L$ lepton number symmetry is
introduced. Table~\ref{tab:ParticleContent2} shows the scalar and
leptonic fields of the model and their representations under the gauge
and global symmetries.

{
\renewcommand{\arraystretch}{1.4}
\begin{table}[tb]
  \centering
  \begin{tabular}{|c|c||ccc|c|}
      \hline
      \textbf{Field} & \textbf{Generations} & \textbf{$\rm SU(3)_c$} & \textbf{$\rm SU(2)_L$} & \textbf{$\rm U(1)_Y$} & \textbf{$\rm U(1)_L$} \\
      \hline
      $\ell_L$ &  3  &  \one  &  \two  &  -1/2  &  1   \\
      $e_R$    &  3  &  \one  &  \one  &  -1    &  1   \\
      $N$      &  3  &  \one  &  \one  &  0     &  $q_N$  \\
      \hline
      $H$      &  1  &  \one  &  \two  &  1/2   &  0   \\
      $\eta$   &  1  &  \one  &  \two  &  1/2   &  $q_\eta$  \\
      $\sigma_1$ &  1  &  \one  &  \one  &  0     &  $q_{\sigma_1}$  \\
      $\sigma_2$ &  1  &  \one  &  \one  &  0     &  $q_{\sigma_2}$  \\  
      $S$      &  1  &  \one  &  \one  &  0     &  $q_S$   \\
      \hline
  \end{tabular}
  \caption{\label{tab:ParticleContent2}
    Lepton and scalar particle content and representations under the gauge and global symmetries in an UV extension of the Scotogenic model with two $\sigma$ fields.}
\end{table}
}

\begin{figure}[tb!]
  \centering
  \includegraphics[width=0.5\linewidth]{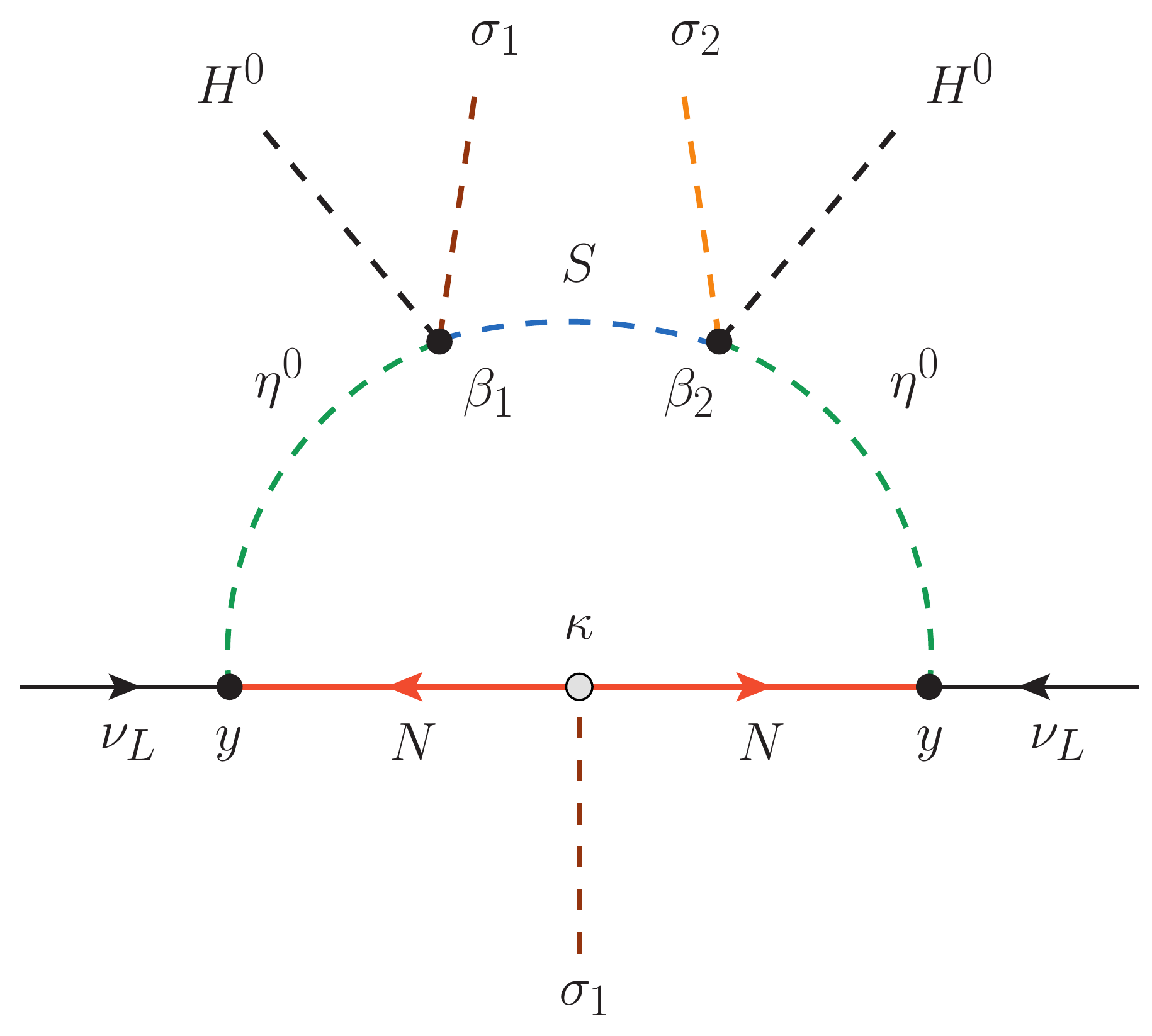}
  \caption{Neutrino mass generation in an extended Scotogenic model
    with two $\sigma$ fields. This Feynman diagram shows the relevant
    gauge eigenstates involved in the 1-loop contribution to neutrino
    masses. In our notation, this is a $\II(1,2)$ model.
    \label{fig:scot2}
    }
\end{figure}

We consider the 1-loop generation of neutrino masses by the diagram in
Fig.~\ref{fig:scot2}. In our notation, this is a $\II(1,2)$ model. For
this mechanism to take place, the operators
\begin{equation}
  \overline{N} \widetilde{\eta}^\dagger \ell_L \quad , \quad \sigma_1 \overline{N}^c N \quad , \quad \sigma_1 H^\dagger S \eta \quad , \quad \sigma_2 H^\dagger S^* \eta
\end{equation}
must be allowed by the symmetries of the model. This restricts the
$\rm U(1)_L$ charges of the fields in the model. In particular, one
can write the following set of equations for them:
\begin{align}
  -q_N + q_\eta + 1 &= 0 \, , \\
  q_{\sigma_1} + 2 \, q_N &= 0 \, , \\
  q_{\sigma_1} + q_S + q_\eta &= 0 \, , \\
  q_{\sigma_2} - q_S + q_\eta &= 0 \, .
\end{align}
They can be solved in terms of $q_N$ to obtain
\begin{align}
  q_\eta &= q_N - 1 \, , \label{eq:qeta} \\
  q_{\sigma_1} &= -2 \, q_N \, , \label{eq:qsig1} \\
  q_{\sigma_2} &= 2 \, , \label{eq:qsig2} \\
  q_S &= q_N + 1 \, . \label{eq:qs}
\end{align}
In addition, we want the operators
\begin{equation}
  \overline{N}^c N \quad , \quad
  \overline{N} \widetilde{H}^\dagger \ell_L \quad , \quad
  \left( H^\dagger \eta \right)^2
\end{equation}
to be forbidden. In order to forbid the first operator, a Majorana
mass term for $N$, we just require $q_N \neq 0$. The second operator
would lead to $\nu_L$-$N$ Dirac mass terms and we can forbid it by
requiring $q_N \neq 1$. Then, Eq.~\eqref{eq:qeta} implies $q_\eta \neq
0$ too. Finally, with these considerations, we choose
\begin{equation}
  q_N = \frac{1}{2} \, ,
\end{equation}
which implies
\begin{equation}
  q_\eta = -\frac{1}{2} \quad , \quad q_S = \frac{3}{2} \quad , \quad q_{\sigma_1} = -1 \quad , \quad q_{\sigma_2} = 2 \, .
\end{equation}
Some comments are in order. First, the diagram in Fig.~\ref{fig:scot2}
has two different $\sigma$ singlets attached to the scalar internal
line, $\sigma_1$ and $\sigma_2$. In principle, one may wonder why we
did not consider the same $\sigma$ singlet in both vertices as
starting point for constructing our model. That would imply $q_S = 0$
and reduce the number of couplings in the model. However, such
construction would lead to an effective operator $(H^\dagger \eta)^2
\sigma^2$ after integrating out the $S$ field. If this operator is
allowed by all symmetries of the model, so is the trilinear
$(H^\dagger \eta) \, \sigma$. We will eventually assume that the
$\sigma$ singlets acquire non-zero VEVs, breaking the original $\rm
U(1)_L$. In the presence of the trilinear $(H^\dagger \eta) \,
\sigma$, this would induce a tadpole for $\eta$, hence breaking the
\z2 parity of the Scotogenic model. This forces us to discard this
possibility and consider different $\sigma_1$ and $\sigma_2$ attached
to the internal scalar line. It also illustrates why models with
$\sigma_A = \sigma_B$ are not compatible with our
requirements. Furthermore, one may consider a third $\sigma_3$ singlet
field coupled to the internal fermion line. While this is possible, we
preferred to choose a charge assignment that allows us to identify
$\sigma_3 \equiv \sigma_1$ and reduce the number of fields in the
model. Finally, once $\sigma_1$ and $\sigma_2$ acquire non-zero VEVs,
the original $\rm U(1)_L$ symmetry will get broken to one of its
$\mathbb{Z}_n$ subgroups. Here $n$ is the GCD of $|q_{\sigma_1}|$ and
$|q_{\sigma_2}|$ after being normalized to become integer numbers,
hence $n = 2$ and the remnant symmetry is \z2.

Once we know the quantum numbers of all the particles in the model, we
can write its Lagrangian. The new Yukawa interactions are given by
\begin{equation}
  \mathcal{L}_{\rm Y} = y \, \overline{N} \, \widetilde{\eta}^\dagger \, \ell_L + \kappa \, \sigma_1 \overline{N}^c N + \hc \, ,
\end{equation}
where $y$ is a general complex $3 \times 3$ matrix and $\kappa$ is a
complex symmetric $3 \times 3$ matrix. The scalar potential of the
model is given by
\begin{equation}
\begin{split}
\mathcal{V}_{\rm UV} &= m_H^2 H^{\dagger} H+m_S^2 S^* S+m_{\sigma_i}^2 \sigma_i^* \sigma_i+ m_{\eta}^2\eta^{\dagger}\eta +\frac{\lambda_1}{2}(H^{\dagger}H)^2+\frac{\lambda_2}{2} (\eta^{\dagger} \eta)^2 \\
&+\frac{\lambda_S}{2}(S^* S)^2+\frac{\lambda_{\sigma_i}}{2}(\sigma^*_{i}\sigma_i)^2 + \lambda_3 (H^{\dagger}H)(\eta^{\dagger}\eta)+\lambda_3^S(H^{\dagger}H)(S^* S) \\
&+\lambda_3^{\sigma_i}(H^{\dagger}H)(\sigma_i^* \sigma_i)+\lambda_3^{\eta S}(\eta^{\dagger}\eta)(S^* S)+\lambda_3^{\eta \sigma_i}(\eta^{\dagger}\eta)(\sigma_i^*\sigma_i) \\
&+\lambda_3^{\sigma\sigma}(\sigma_1^*\sigma_1)(\sigma_2^*\sigma_2)+\lambda_3^{\sigma_i S}(\sigma_i^*\sigma_i)(S^* S)+\lambda_4 (H^{\dagger}\eta)(\eta^{\dagger}H) \\
&+\left[\beta_1 (\sigma_1 H^{\dagger} S \eta)+\beta_2 (\sigma_2 H^{\dagger} S^\dagger \eta)+\frac{\mu}{\sqrt{2}} (\sigma_2 \sigma_1 \sigma_1)+\lambda_0 (S S \sigma_1 \sigma_2^*)+ \hc \right] \, ,
\end{split}\label{UUV2}
\end{equation}
where we sum over $i=1,2$. Here $\mu$ is a trilinear parameter with
dimensions of mass while $m_H^2$, $m_\eta^2$ and $m_{\sigma_i}^2$ have
dimensions of mass$^2$. Other Lagrangian terms are allowed by the
gauge symmetries of the model but forbidden by $\rm U(1)_L$.

\subsection{Effective theory}

In the following we will assume that $m_S$ is much larger than any
other energy scale in the model and integrate out the heavy scalar
$S$. If we do this at tree-level, the effective scalar potential at
low energies can be written as
\begin{equation}
\begin{split}
\mathcal{V}_{\rm IR}&=m_H^2(H^{\dagger}H)+m_{\eta}^2 (\eta^{\dagger}\eta)+m_{\sigma_i}^2(\sigma_i^* \sigma_i)+\frac{\lambda_1}{2}(H^{\dagger}H)^2+\frac{\lambda_2}{2}(\eta^{\dagger}\eta)^2 +\frac{\lambda_{\sigma_i}}{2}(\sigma_i^*\sigma_i)^2\\
&+ \lambda_3 (H^{\dagger}H)(\eta^{\dagger}\eta) +\lambda_3^{\sigma_i}(H^{\dagger}H)(\sigma_i^*\sigma_i)+\lambda_3^{\eta \sigma_i}(\eta^{\dagger}\eta)(\sigma^*_i\sigma_i)+\lambda_3^{\sigma\sigma
}(\sigma_1^*\sigma_1)(\sigma_2^*\sigma_2)\\
&+\left[\lambda_4-\frac{|\beta_i|^2}{m_S^2}(\sigma_i^*\sigma_i)\right](H^{\dagger}\eta)(\eta^{\dagger}H)\\
&+\left[\frac{\mu}{\sqrt{2}} (\sigma_2 \sigma_1\sigma_1)-\frac{\beta_1\beta_2}{m_S^2}\sigma_1\sigma_2(H^{\dagger}\eta)^2+ \hc \right]+\mathcal{O}\left(\frac{1}{m_S^4}\right) \, .
\end{split}\label{eq:UIR2}
\end{equation}
Now, we decompose the neutral fields $H^0$ and $\sigma_{1,2}$ as
\begin{equation}
  H^0=\frac{1}{\sqrt{2}}(v_H+\phi+i \, A) \, , \quad
  \sigma_i=\frac{1}{\sqrt{2}}(v_{\sigma_i}+\rho_i+i \, J_i) \, , 
\label{H0Sigma}
\end{equation}
where we defined $\frac{v_H}{\sqrt{2}}$ and
$\frac{v_{\sigma_i}}{\sqrt{2}}$ as the VEVs of the corresponding
fields. After this, we can compute the tadpole equation resulting from the
effective potential in Eq.~\eqref{eq:UIR2}, evaluated at the VEVs of
each scalar field. The non-trivial tadpole equations are
\begin{align}
\left.\frac{d \mathcal{V}_{\rm IR}}{d H^0}\right|_{\langle H^0,\sigma_i\rangle=\{\frac{v_{H}}{\sqrt{2}},\frac{v_{\sigma_i}}{\sqrt{2}}\}}&= \frac{v_{H}}{\sqrt{2}}\left(m_{H}^2+\lambda_1 \frac{v_{H}^{2}}{2} +\lambda_3^{\sigma_1}\frac{v_{\sigma_1}^{2}}{2}+\lambda_3^{\sigma_2}\frac{v_{\sigma_2}^{2}}{2}\right)=0\label{tp1},\\
\left.\frac{d \mathcal{V}_{\rm IR}}{d \sigma_1}\right|_{\langle H^0,\sigma_i\rangle=\{\frac{v_{H}}{\sqrt{2}},\frac{v_{\sigma_i}}{\sqrt{2}}\}}&=\frac{v_{\sigma_1}}{\sqrt{2}}\left(m_{\sigma_1}^2+\mu \,  v_{\sigma_2}+\lambda_3^{\sigma_1}\frac{v_{H} ^2}{2}+\lambda_{\sigma_1}\frac{v_{\sigma_1}^2}{2}+\lambda_3^{\sigma\sigma}\frac{v_{\sigma_2}^2}{2}\right)=0\label{tp2},\\
\left.\frac{d \mathcal{V}_{\rm IR}}{d \sigma_2}\right|_{\langle H^0,\sigma_i\rangle=\{\frac{v_{H}}{\sqrt{2}},\frac{v_{\sigma_i}}{\sqrt{2}}\}}&=\frac{v_{\sigma_2}}{\sqrt{2}}\left(m_{\sigma_2}^2+\mu\frac{v_{\sigma_1}^2}{2 v_{\sigma_2}}+\lambda_3^{\sigma_2}\frac{v_{H} ^2}{2}+\lambda_{\sigma_2}\frac{v_{\sigma_2}^2}{2}+\lambda_3^{\sigma\sigma}\frac{v_{\sigma_1}^2}{2}\right)=0\label{tp3}.
\end{align}
As already explained, as a result of $\sigma_i$ acquiring a VEV,
lepton number gets spontaneously broken, leaving a discrete \z2
symmetry, under which all the particles in the model are even except
for $N$ and $\eta$, which are odd. Another important consequence of
the spontaneous breaking of lepton number is the generation of the
$(H^{\dagger}\eta)^2$ operator, with a naturally suppressed
$\lambda_5$ coupling due to the $1/m_S^2$ factor. One finds
\begin{equation}
\frac{\lambda_5}{2}=-\frac{v_{\sigma_1}v_{\sigma_2}\beta_1\beta_2}{2m_S^2}\ll 1 \, ,\label{eq:lambda52}
\end{equation}
where $\beta_i$ are dimensionless parameters of the UV theory and
$v_{\sigma_i}\ll m_S$ by construction. This expression clearly
corresponds to a $\II(1,2)$ model, following the classification of
Sec.~\ref{sec:clasif}. We now consider the scalar spectrum of the
model. We will assume that CP is conserved in the scalar sector, just
for the sake of simplicity. In this case, the spectrum contains three
CP-even and three CP-odd gauge eigenstates. In the bases
$\{\phi,\rho_1,\rho_2\}$ and $\{A,J_1,J_2 \}$, their mass matrices are
given by
\begin{equation}
\mathcal{M}_R^2=\left(\begin{array}{ccc}
\lambda_1 v_{H}^2&\lambda_3^{\sigma_1} v_{H} v_{\sigma_1} & \lambda_3^{\sigma_2} v_{H} v_{\sigma_2}\\
\lambda_3^{\sigma_1} v_{H} v_{\sigma_1}&\lambda_{\sigma_1} v_{\sigma_1}^2&v_{\sigma_1}(\mu+\lambda_3^{\sigma\sigma}v_{\sigma_2})\\
\lambda_3^{\sigma_2}v_{H} v_{\sigma_2}&v_{\sigma_1}(\mu+\lambda_3^{\sigma\sigma}v_{\sigma_2})&\lambda_2 v_{\sigma_2}^2-\frac{\mu v_{\sigma_1}^2}{2v_{\sigma_2}}\\
\end{array}\right) \label{CPeven}
\end{equation}
and
\begin{equation}
\mathcal{M}_I^2=\left(\begin{array}{ccc}
0&0&0\\
0&-2\mu v_{\sigma_2}&-\mu v_{\sigma_1}\\
0&-\mu v_{\sigma_1}&-\frac{\mu v_{\sigma_1}^2}{2v_{\sigma_2}}
\end{array}\right) \, , \label{CPodd}
\end{equation}
respectively. The tadpole equations~\eqref{tp1}-\eqref{tp3} were used
in the derivation of Eqs.~\eqref{CPeven} and \eqref{CPodd}. The
CP-even and CP-odd physical mass eigenstates can be written as linear
combinations of $\{\phi,\rho_1,\rho_2\}$ and $\{A,J_1,J_2\}$,
respectively, obtained after the diagonalization of the matrices
$\mathcal{M}_R^2$ and $\mathcal{M}_I^2$. Out of the three CP-even mass
eigenstates, one can be identified with the Higgs boson, with mass
$m_h \simeq 125$ GeV, discovered at the LHC. In addition, two massive
CP-even scalar fields exist. In what concerns the CP-odd mass
eigenstates, their mass matrix in Eq.~\eqref{CPodd} can be readily
diagonalized as $V_I^T \, \mathcal{M}_I^2 \, V_I =
\widehat{\mathcal{M}}_I^2$, where
\begin{equation} \label{eq:thetadef}
  V_I = \left(\begin{array}{ccc}
1&0&0\\
0& \cos \theta & - \sin \theta \\
0& \sin \theta & \cos \theta
\end{array}\right)
\end{equation}
is a unitary matrix and $\widehat{\mathcal{M}}_I^2$ is a diagonal
matrix. One obtains
\begin{equation}
\widehat{\mathcal{M}}_I^2=\left(\begin{array}{ccc}
0&0&0\\
0&0&0\\
0&0&-\frac{\mu(v_{\sigma_1}^2+4v_{\sigma_2}^2)}{2v_{\sigma_2}}
\end{array}\right) \, ,\label{CPoddD}
\end{equation}
thus leading to two massless pseudoscalar bosons. The first one is the
Goldstone boson that becomes the longitudinal component of the $Z$
boson ($A$), while the second one (a linear combination of fields
$J_1$ and $J_2$) is associated to the spontaneous breaking of $\rm
U(1)_L$ and is the so-called majoron, denoted as $J$. The $J_1-J_2$
mixing angle is given by
\begin{equation}
\tan(2\theta)=\frac{2 \, (\mathcal{M}_I^2)_{23}}{(\mathcal{M}_I^2)_{22}-(\mathcal{M}_I^2)_{33}}=\frac{4 v_{\sigma_1}v_{\sigma_2}}{4v_{\sigma_2}^2-v_{\sigma_1}^2} \, .\label{MajoronMixing}
\end{equation}
We finally turn our attention to the \z2-odd scalars and decompose the
neutral field $\eta^0$ as
\begin{equation}
\eta^0=\frac{1}{\sqrt{2}}(\eta_R+ i \, \eta_I) \, .
\end{equation}
The mass of the charged $\eta^+$ and the neutral $\eta_{R,I}$ fields
are given by
\begin{align}
m_{\eta^{+}}^2&=m_{\eta}^2+\frac{v_H^2}{2} \lambda_3^{\rm eff} \, ,\label{etapmass}\\
m_{\eta_R}^2&=m_{\eta}^2+\frac{v_{H}^{2}}{2}\left(\lambda_3^{\rm eff} +\lambda_4^{\rm eff}-\frac{\beta_1\beta_2  v_{\sigma_1} v_{\sigma_2}}{m_S^2}\right) \, ,\label{etaRmass} \\
m_{\eta_I}^2&=m_{\eta}^2+\frac{v_{H}^2}{2}\left(\lambda_3^{\rm eff} +\lambda_4^{\rm eff}+\frac{\beta_1\beta_2  v_{\sigma_1} v_{\sigma_2}}{m_S^2}\right) \, ,\label{etaImass}
\end{align}
where we defined
\begin{align}
\lambda_3^{\rm eff} &\equiv \lambda_3 +\lambda_3^{\eta\sigma_1} \frac{v_{\sigma_1}^2}{v_H^2}+\lambda_3^{\eta\sigma_2} \frac{v_{\sigma_2}^2}{v_H^2} \, \\
\lambda_4^{\rm eff} &\equiv \lambda_4 -\frac{\beta_1^2  v_{\sigma_1}^2}{2m_S^2}-\frac{\beta_2^2 v_{\sigma_2}^2}{2 m_S^2} \, .
\end{align}
As in the Scotogenic model, the mass difference between $\eta_R$ and
$\eta_I$ is proportional to the $\lambda_5$ coupling:
\begin{equation}
m_{\eta_R}^2-m_{\eta_I}^2=-\frac{v_{\sigma_1}v_{\sigma_2}\beta_1\beta_2}{m_S^2} v_H ^2=\lambda_5 v_H^2 \, .
\end{equation}
Finally, the breaking of $\rm U(1)_L$ also induces a Majorana mass
term for the $N$ singlets, with $M_N = \sqrt{2} \, \kappa \,
v_{\sigma_1}$. This leads to Majorana neutrino masses at 1-loop, as
shown in Fig.~\ref{fig:scot2}. The resulting neutrino mass matrix is
given by Eq.~\eqref{eq:numass}, with the effective $\lambda_5$ of
Eq.~\eqref{eq:lambda52}. Furthermore, contrary to the minimal
Scotogenic model, this UV extension induces a 1-loop interaction
between the majoron and a pair of charged leptons. This enriches the
phenomenology of the model, as we discuss in the next Section.

\section{Phenomenology}
\label{sec:pheno}

All UV scenarios discussed in our classification of
Sec.~\ref{sec:clasif} and illustrated with the two examples of
Secs.~\ref{sec:model1} and \ref{sec:model2} share some common
features. They are characterized at low energies by a Scotogenic model
extended with a massless pseudoscalar, the majoron $J$, and one or
several massive scalars and pseudoscalars. While some phenomenological
implications may be specific to particular models, there are also some
general expectations that we may highlight.

\subsection{Majoron coupling to charged leptons}
\label{subsec:Jcoup}

The presence of a massless majoron dramatically affects the
phenomenology of this class of models. In fact, models including a
majoron are strongly constrained by a variety of experimental limits,
such as those originated by the majoron coupling to a pair of charged
leptons. The relevance of these limits depends on the flavor structure
of the couplings~\cite{Sun:2021jpw}, which necessarily depends on the
specific model. Stringent constraints exist for both flavor-conserving
and flavor-violating couplings. Let us write the majoron interaction
with charged leptons as~\cite{Escribano:2020wua},
\begin{equation} \label{eq:llJ}
\mathcal{L}_{\ell \ell J} = J \, \bar{\ell}_\beta \left( S_L^{\beta \alpha} \, P_L + S_R^{\beta \alpha} \, P_R \right)\ell_{\alpha} + \hc \, .
\end{equation}
Here $\ell_{\alpha,\beta}$ are the charged leptons with flavors
$\alpha$ and $\beta$, while $P_{L,R}$ are the usual chiral
projectors. We consider all flavor combinations for the $S_{L,R}$
couplings: $\beta\alpha = \left\{
ee,\mu\mu,\tau\tau,e\mu,e\tau,\mu\tau\right\}$. Due to the
pseudoscalar nature of majorons, the diagonal $S^{\beta \beta} =
S_L^{\beta \beta} + S_R^{\beta \beta \ast}$ couplings are purely
imaginary. They receive strong constraints from astrophysical
observations, due to the cooling effects induced by the majoron in
dense astrophysical media. Flavor off-diagonal couplings are
constrained by the null searches of lepton flavor violation in
processes involving charged leptons. In particular, searches for
$\ell_\alpha \to \ell_\beta \, J$ can be used to set bounds on the
combinations
\begin{equation}
|S^{\beta\alpha}|=\left( \left| S_L^{\beta\alpha} \right|^2 + \left| S_R^{\beta\alpha} \right|^2 \right)^{1/2} \, .
\end{equation}
A compilation of the current limits on the majoron couplings to
charged leptons can be found in Table~\ref{tab:boundsJll}.

{
\renewcommand{\arraystretch}{1.4}
\begin{table}[tb]
  \centering
  \begin{tabular}{|c|c|c|}
      \hline
      \textbf{Coupling} & \textbf{Upper limit} & \textbf{References} \\
      \hline
      $\text{Im} \, S^{e e}$ & $2.1 \times 10^{-13}$ & \cite{Calibbi:2020jvd} \\
      $\text{Im} \, S^{\mu \mu}$ & $2.1 \times 10^{-9}$ & \cite{Croon:2020lrf} \\
      \hline
      $|S^{e\mu}|$ & $5.3 \times 10^{-11}$ & \cite{Escribano:2020wua} \\
      $|S^{e\tau}|$ & $5.9 \times 10^{-7}$ & \cite{Escribano:2020wua} \\
      $|S^{\mu\tau}|$ & $7.6 \times 10^{-7}$ & \cite{Escribano:2020wua} \\      
      \hline
  \end{tabular}
  \caption{
    Current limits on the majoron couplings to charged leptons. The
    limit on $\text{Im} \, S^{e e}$ is at 90\%
    C.L.~\cite{Calibbi:2020jvd}. The limit on $\text{Im} \, S^{\mu
      \mu}$ has been obtained by performing a simulation of the
    supernova SN1987A~\cite{Croon:2020lrf}. An alternative and more
    stringent limit $\text{Im} \, S^{\mu \mu} < 2.1 \times 10^{-10}$
    can be derived with more aggressive assumptions in the simulation.
  \label{tab:boundsJll}}
\end{table}
}

\begin{figure}[tb!]
  \centering
  \includegraphics[width=0.5\linewidth]{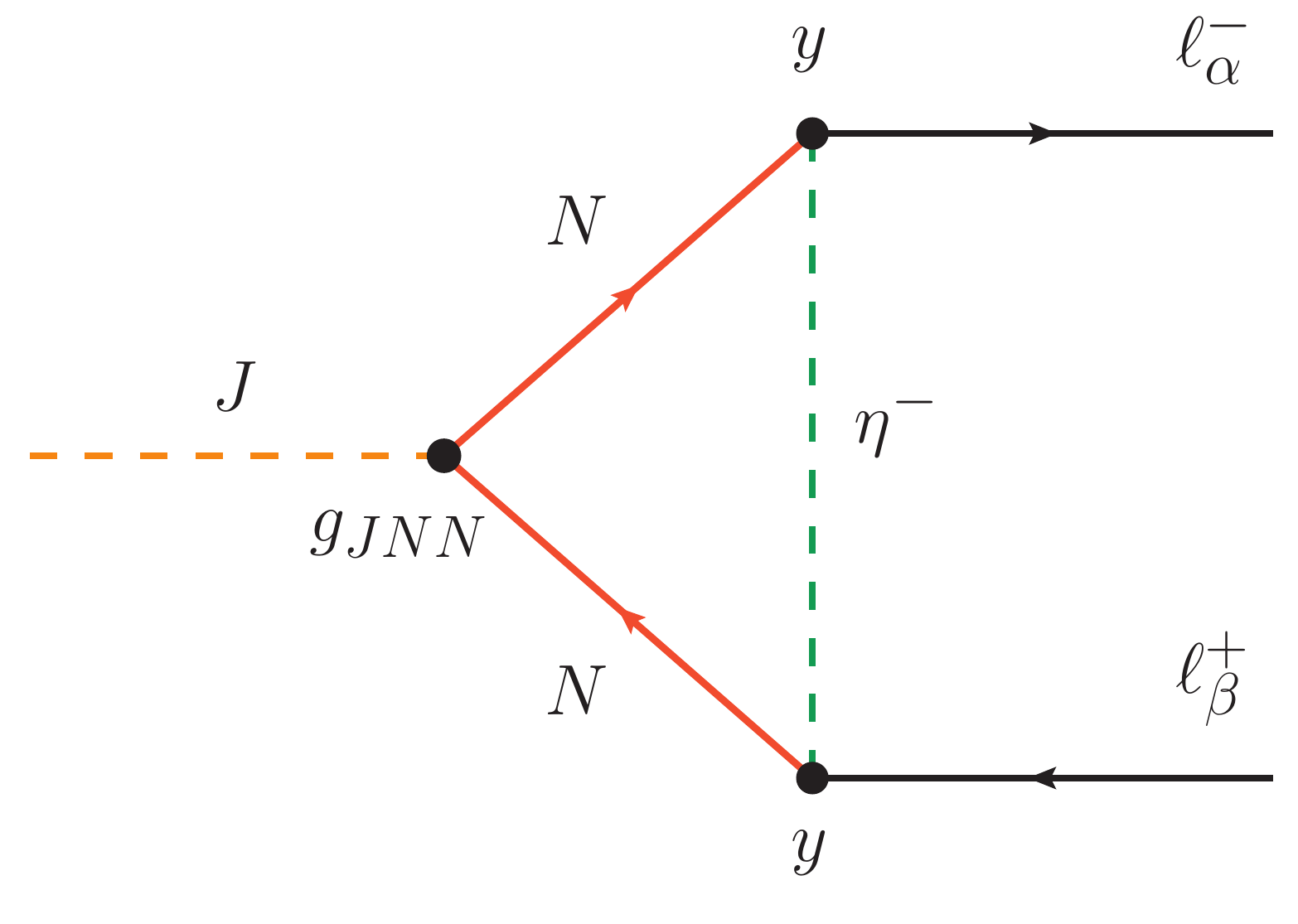}
  \caption{1-loop generation of the majoron coupling to a pair of
    charged leptons in the Scotogenic scenarios discussed in this
    work.
    \label{fig:Jll}
    }
\end{figure}

While in some scenarios the majoron couplings to charged leptons
appear at tree-level~\cite{Hirsch:2009ee,Escribano:2021uhf}, in many
cases the leading order contribution is induced at the 1-loop
level. For instance, this is the case of the popular type-I seesaw
with spontaneous lepton number
violation~\cite{Chikashige:1980ui,Pilaftsis:1993af,Heeck:2019guh}. Similarly,
in the Scotogenic scenarios discussed in this paper, the majoron
coupling to charged leptons is also induced at
1-loop~\cite{Babu:2007sm,Escribano:2021ymx} by the Feynman diagram in
Fig.~\ref{fig:Jll}. Here $g_{JNN}$ is the $J-N-N$ coupling, which
depends on the specific model. It is given by
\begin{equation}
  g_{JNN} = \left\{ \begin{array}{cl}
    i \frac{\kappa}{\sqrt{2}} & \text{in models with one $\sigma$ singlet} \\
    i \frac{\kappa}{\sqrt{2}} \, \cos \theta & \text{in models with two $\sigma$ singlets} \end{array} \right. \quad ,
\end{equation}
where the mixing angle $\theta$ is defined in Eq.~\eqref{eq:thetadef}.
The prefactor $\cos \theta$ in models with two $\sigma$ singlets is
due to the fact that only $\sigma_1$ has a coupling to $\overline{N}^c
N$. No other contributions to the majoron coupling to charged leptons
exist at 1-loop. One may wonder about a Feynman diagram with two
scalar lines in the loop, induced by a $J \, \eta^+ \eta^-$
coupling. However, this contribution vanishes exactly. The reason is
the pseudoscalar nature of the majoron. The $J \bar \ell_\alpha
\ell_\alpha$ vertex must be proportional to $\gamma_5$, but the
Lorentz structure of this contribution does not generate such
pseudoscalar coupling.~\footnote{We also note that the $J \, \eta^+
\eta^-$ coupling is absent in many models, since Lagrangian terms like
$\sigma |\eta|^2$ or $\sigma^2 |\eta|^2$ are forbidden by lepton
number. Only in models with two $\sigma$ fields one may have a term of
the form $\sigma_1 \sigma_2 |\eta|^2$ (when $q_{\sigma_1} = -
q_{\sigma_2}$) leading to a $J \, \eta^+ \eta^-$ interaction vertex
after symmetry breaking. However, as explained in the text, even when
this term is present, the associated 1-loop contribution to the
majoron coupling to a pair of charged leptons vanishes exactly due to
the pseudoscalar nature of the majoron.} Also, diagrams with gauge
bosons vanish due to the pure singlet nature of $N$. Therefore, one
can find the $S_{L,R}$ couplings introduced in Eq.~\eqref{eq:llJ} by
direct computation of the diagram in Fig.~\ref{fig:Jll}. The result
can be written as~\cite{Babu:2007sm,Escribano:2021ymx}
\begin{align}
  & S_L^{\beta \alpha} = - \frac{m_{\ell_\beta}}{8\pi^2} \left(y^\dagger g_{JNN} \, \Gamma \, y \right)_{\beta \alpha} \, , \\
  & S_R^{\beta \alpha} = \frac{m_{\ell_\alpha}}{8\pi^2} \left(y^\dagger g_{JNN} \, \Gamma \, y \right)_{\beta \alpha} \, ,
\end{align}
for the non-diagonal couplings and
\begin{equation}
  S^{\beta \beta} = - \frac{m_{\ell_\beta}}{8\pi^2} \left(y^\dagger g_{JNN} \, \Gamma \, y \right)_{\beta \beta} \, ,
\end{equation}
for the diagonal ones. Here $m_{\ell_{\beta}} = \{m_e,m_\mu,m_\tau\}$
and we have defined
\begin{equation}
  \Gamma_{m n} = \frac{M_{N_n}}{\left( M_{N_n}^2 - m^2_{\eta^+} \right)^2} \left( M_{N_n}^2 - m^2_{\eta^+} + m^2_{\eta^+} \, \log \frac{m^2_{\eta^+}}{M_{N_n}^2} \right) \delta_{mn} \, .
\end{equation}
\begin{figure}[t]
  \centering
  \includegraphics[width=0.48\linewidth]{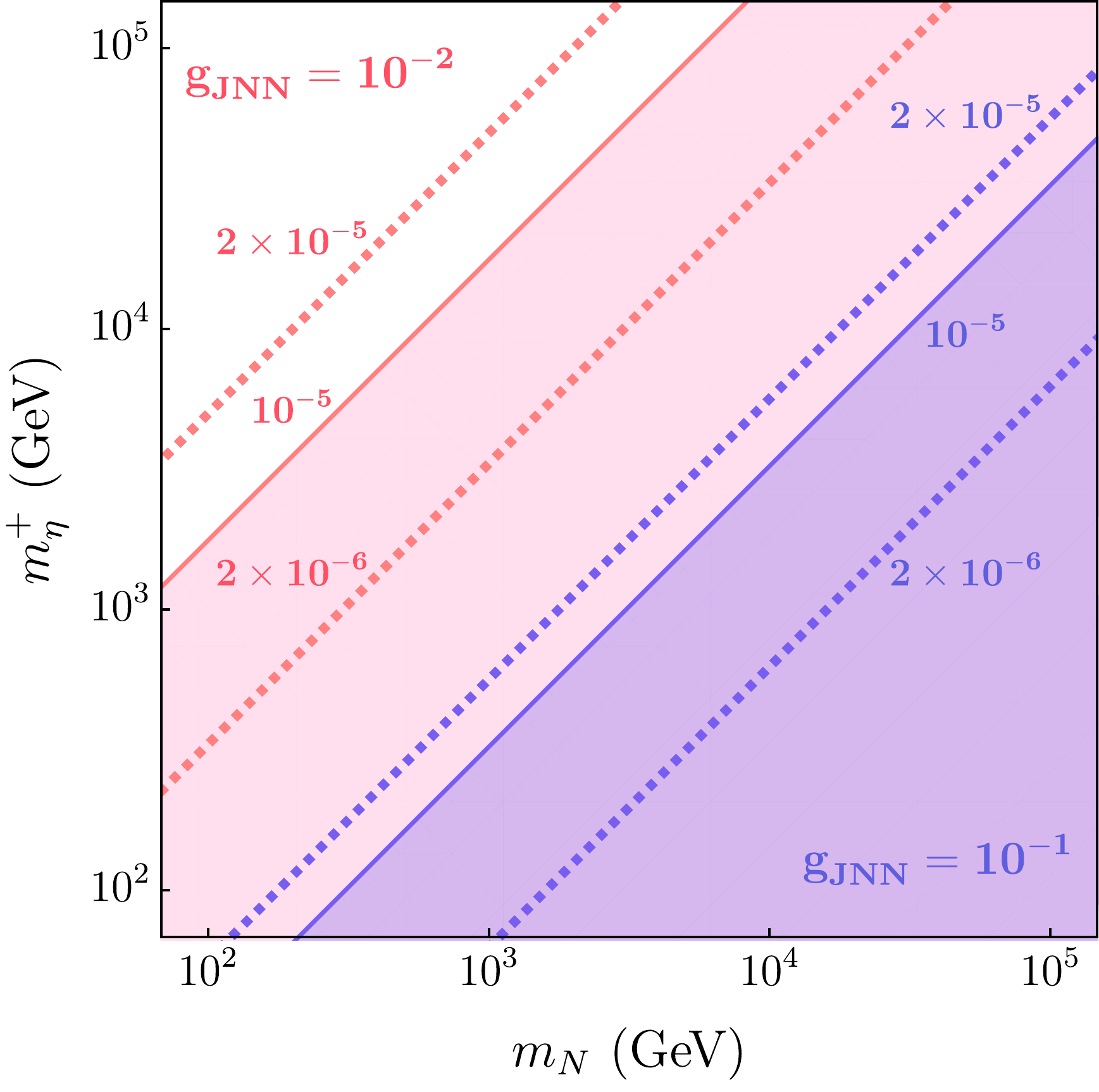}
  \hspace*{0.2cm}
  \includegraphics[width=0.48\linewidth]{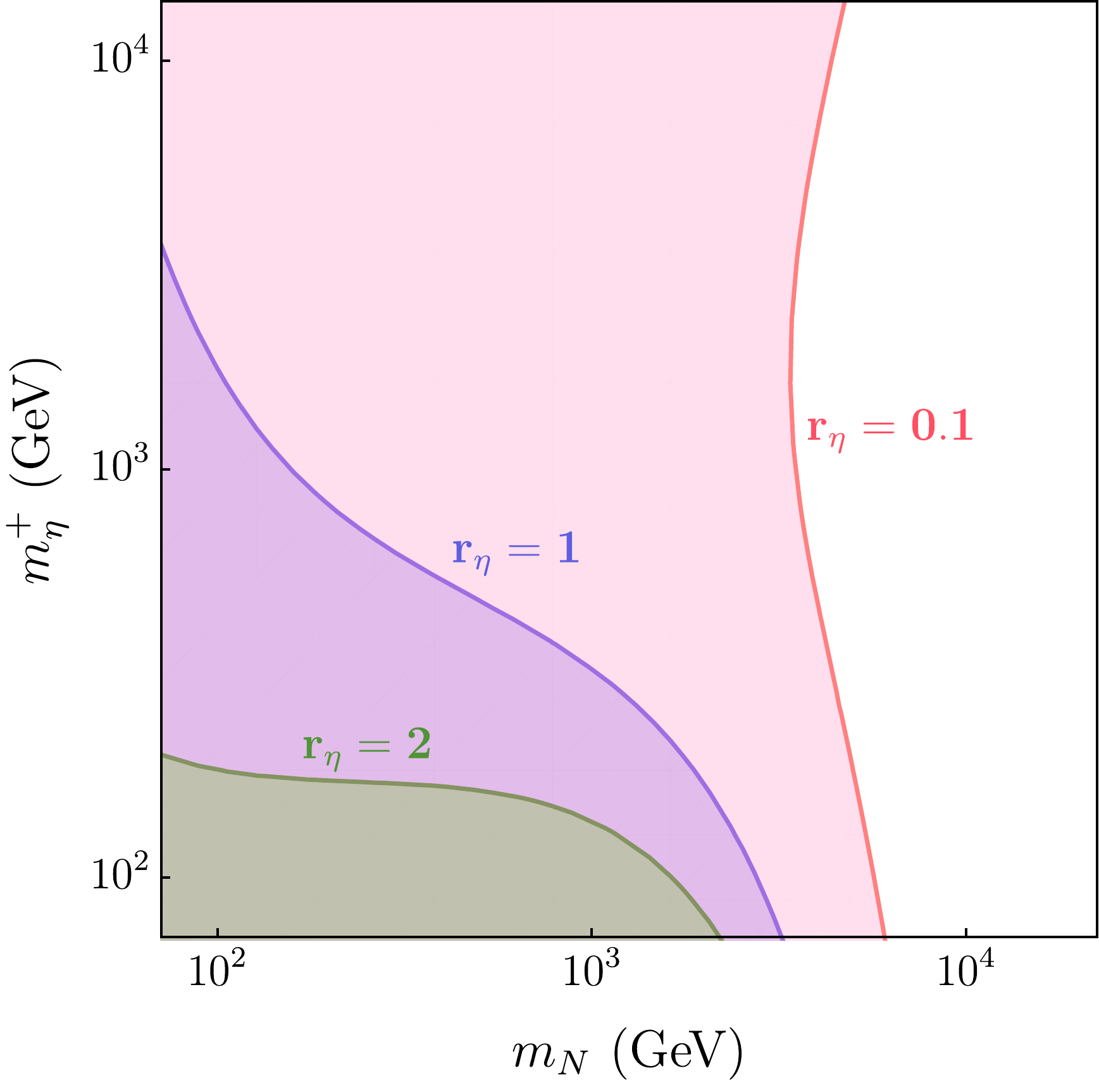}
  \caption{
Contours of $\text{BR}\left(\mu \to e J\right)$ in the $\left(M_N,
m_{\eta^+} \right)$ plane. The colored regions correspond to the
regions allowed by the current experimental bound on the branching
ratio. On the left, $g_{JNN}$ has been fixed to $10^{-1}$ (blue) and
to $10^{-2}$ (pink), while $r_\eta = 1$ has been used. On the right,
the coupling $g_{JNN}$ was not fixed and three different values of the
$r_\eta$ ratio have been considered, $0.1$ (pink), $1$ (blue) and $2$
(green).
  \label{fig:mueJ}
  }
\end{figure}
We can now study how the bounds on these couplings restrict the
parameter space of the models considered in our classification. In
particular, in the following we focus on the 2-body decay $\mu \to e
J$, for which
\begin{equation}
  \text{BR} \left( \mu \rightarrow e J \right) = \frac{m_\mu}{32 \, \pi \, \Gamma_\mu} \left( \left| S_L^{e \mu} \right|^2 + \left| S_R^{e \mu} \right|^2 \right) \, ,
\end{equation}
where $\Gamma_\mu \approx 3 \times 10^{-19}$ GeV is the total decay
width of the muon. We used a Casas-Ibarra
parametrization~\cite{Casas:2001sr} properly adapted to the Scotogenic
model~\cite{Toma:2013zsa, Cordero-Carrion:2018xre,
  Cordero-Carrion:2019qtu} and the best-fit values obtained in the
global fit~\cite{deSalas:2020pgw} to neutrino oscillation data in
order to express the Yukawa matrix $y$ in terms of experimentally
measured quantities. We assumed that the three singlet fermions are
degenerate, that is $M_{N_1} = M_{N_2} = M_{N_3} = M_{N}$ and we fixed
$\lambda_5 = 5 \times 10^{-8}$. Notice that lower values of this
parameter would imply larger values of the Yukawas, thus further
restricting the parameter space of the model. It also proves
convenient to define
\begin{equation}
  r_\eta = \frac{m_0}{m_{\eta^+}} \, .
\end{equation}
Our results are shown in Fig.~\ref{fig:mueJ}. On the left-hand side we
fixed the coupling $g_{JNN}$ to $10^{-1}$ (blue), and to $10^{-2}$ (pink),
and we considered $r_\eta = 1$ in both scenarios. The colored regions
correspond to regions allowed by the experimental bound on the $\mu
\to e J$ decay, which implies $\text{BR} \left( \mu \to e J \right) <
10^{-5}$~\cite{Hirsch:2009ee}. As expected, the larger the $J-N-N$
coupling is, the smaller the allowed region of the parameter space
becomes. We also find that light Scotogenic states can be made
compatible with the $\mu \to e J$ bound. This can be easily understood
by inspecting the non-trivial relation between the masses $m_{\eta^+}$
and $M_N$ and the Yukawa couplings $y$. Under the assumptions
mentioned above one finds
\begin{equation}
  S_{L,R} \propto g_{JNN} \Gamma_{ii} \left( y^\dagger y \right)_{1 2} \, ,
\end{equation}
where $\Gamma_{ii}$ is any of the diagonal entries of $\Gamma$, given by
\begin{equation}
  \Gamma_{ii} \propto M_N \frac{M_{N}^2 - m^2_{\eta^+} + m^2_{\eta^+} \, \log \frac{m^2_{\eta^+}}{M_{N}^2}}{\left( M_{N}^2 - m^2_{\eta^+} \right)^2} \, .
\end{equation}
Eq.~\eqref{eq:numass} implies that the Yukawa product $\left(
y^\dagger y \right)_{1 2}$ is proportional to
\begin{equation}
  \left( y^\dagger y \right)_{1 2} \propto \frac{1}{M_N} \frac{\left( M_{N}^2 - m^2_{0} \right)^2}{M_{N}^2 - m^2_{0} + M^2_{N} \, \log \frac{m^2_{0}}{M_{N}^2}} \, .
\end{equation}
Therefore, in the limit $r_\eta = 1$ one finds
\begin{equation}
  S_{L,R} \propto g_{JNN} \frac{M_N^2 - m_{\eta^+}^2 + m_{\eta^+}^2 \log \left( \frac{m_{\eta^+}^2}{M_N^2} \right)}{M_N^2 - m_{\eta^+}^2 + M_{N}^2 \log \left( \frac{m_{\eta^+}^2}{M_N^2} \right)}  \, .
\end{equation}
For a fixed $g_{JNN}$ value two possibilities arise: (i) if we fix
$m_{\eta^+}$, the $\Gamma_{ii} \, \left( y^\dagger y \right)_{1 2}$ combination
decreases if $M_N$ increases, and (ii) if we fix $M_N$, the $\Gamma_{ii} \, \left(
y^\dagger y \right)_{1 2}$ combination increases if $m_{\eta^+}$
increases. Essentially, the involved couplings strongly depend on
$m_{\eta^+}$ and $M_N$ and this dependence may lead to an apparent
non-decoupling behavior that explains the results for the $\mu \to e
J$ branching ratio observed in Fig.~\ref{fig:mueJ}. Finally, the
right-hand side of this figure provides complementary
information. Here we considered $g_{JNN} = i \frac{\kappa}{\sqrt{2}} =
i \frac{M_N}{2 \, v_{\sigma}}$ and fixed $v_\sigma = 5$ TeV. Since the
$g_{JNN}$ coupling grows with $M_N$, for each $m_{\eta^+}$ there is a
maximum value of $M_N$ for which $\text{BR}\left( \mu \to e J \right)
< 10^{-5}$. This can be clearly seen in our results.

\subsection{Collider signatures}
\label{subsec:collider}

Since the spontaneous breaking of $\rm U(1)_L$ requires the
introduction of additional scalar multiplets, all models in our
classification have extended scalar sectors containing several states
besides the ones in the Scotogenic model. This can be used to probe
them at colliders.

One of the CP-even scalars, presumably the lightest, is to be
identified with the $125$ GeV state discovered at the LHC. The
production cross-section and decay rates of this state, denoted
generally as $h$, must agree with the values measured by the ATLAS and
CMS collaborations. Since these are very close to those predicted for
a pure SM Higgs, $h \approx \text{Re}(H^0)$ is generally required. In
particular, mixings with the $\sigma$ states are strongly constrained,
since they would affect its decay rates in a twofold way. First, the
$\sigma$ states do not couple to the SM gauge bosons or to
quarks. Thus, any mixing would induce a universal reduction of the $h$
partial decay widths into these states. And second, $h$ can have
additional decay modes. It can decay invisibly to a pair of singlet
fermions ($h \to N_1 N_1$) or to pair of majorons ($h \to J J$). The
former can only take place if $m_{N_1} \leq m_h/2$. In contrast,
since the majoron is massless, the latter is always kinematically
available. We can write the interaction Lagrangian of $h$ with a pair
of majorons as $\mathcal{L}_{h J J} = \frac{1}{2} \, g_{h J J} \, h \,
J^2$, where $g_{h J J}$ is a dimensionful coupling that depends on the
specific model. This interaction induces the invisible decay $h \to J
J$, with the decay width given by
\begin{equation}
  \Gamma(h \to J J) = \frac{g_{h J J}^2}{32 \, \pi \, m_h} \, .
\end{equation}
If we assume a total Higgs decay width in agreement with the SM
expectation, $\Gamma_h \approx \Gamma_h^{\rm SM} = 4.1$
MeV~\cite{LHCHiggsCrossSectionWorkingGroup:2016ypw}, the bound on the
invisible Higgs branching ratio $\text{BR}(h \to J J) < 0.19$ at
$95\%$ C.L.~\cite{CMS:2018yfx}, implies $g_{h J J} < 3.1$ GeV. This
translates into constraints on the parameters of the scalar potential
of the model, which are encoded in $g_{h J J}$. For instance, in the
model discussed in~\cite{Escribano:2021ymx}, this implies that the
coefficient of the $(H^{\dagger}H)(\sigma^* \sigma)$ operator must be
$\lesssim 10^{-2}$. We note, however, that stronger constraints can be
derived by combining invisible and visible channels, as recently
pointed out in~\cite{Biekotter:2022ckj}.

Finally, all models in our classification also contain additional
heavy states. They can also be searched for at colliders. Their
production cross-sections and decay models strongly depend on the
specific realization of our setup and, more specifically, on their
gauge composition. If they have sizable doublet components, they can
in principle be produced at high rates at the LHC via Drell–Yan
processes. In contrast, heavy scalars with a dominant component in
the singlet direction have very suppressed production cross-sections
at the LHC. Due to the constraints discussed above, which imply
suppressed mixing between the SM Higgs doublet and the $\sigma$
states, this is the most likely scenario in all models discussed in
our classification.

\subsection{Dark matter}
\label{subsec:dm}

In all UV models studied in this paper, a remnant \z2 symmetry is
obtained as a result of the spontaneous breaking of lepton
number. This is the Scotogenic \z2 parity, under which only the usual
Scotogenic states $N$ and $\eta$ are charged. The conservation of \z2
implies that the lightest of them is completely stable and, in
principle, a valid DM candidate. Both options have been widely
studied in the literature. In the case of a scalar candidate, the DM
phenomenology resembles that of the Inert Doublet
model~\cite{Deshpande:1977rw,Barbieri:2006dq,LopezHonorez:2006gr,LopezHonorez:2010eeh,Diaz:2015pyv},
with the DM production in the early Universe set by gauge
interactions. In contrast, the case of a fermion candidate typically
requires large Yukawa couplings. This leads to tension with bounds
from lepton flavor violation~\cite{Toma:2013zsa}, although the
observed DM relic density can be
achieved~\cite{Kubo:2006yx,AristizabalSierra:2008cnr,Suematsu:2009ww,Adulpravitchai:2009gi,Vicente:2014wga}.

The low energy theories resulting from our UV extended models do not
correspond \textit{exactly} to the original Scotogenic model. As
explained above and illustrated in Secs.~\ref{sec:model1} and
\ref{sec:model2}, additional scalar states are present: the massless
majoron and one or several massive scalars. These new degrees of
freedom couple to the \z2-odd states and may affect the resulting DM
phenomenology, which may have some differences with respect to the one
in the original Scotogenic scenario. This has recently been studied
in~\cite{Bonilla:2019ipe,DeRomeri:2022cem} for the case of fermion
DM. The main conclusion from these works is that the new scalar states
open up new regions in parameter space in which the DM relic density
can match the observed value. In particular, annihilations become very
efficient when the mass of the DM candidate, $m_{N_1}$, is about half
of the mass of a new scalar state. This implies that one can find the
correct DM abundance for any value of $m_{N_1}$ without resorting to
coannihilations, in contrast to the original Scotogenic model. These
models are also expected to have a rich phenomenology at direct and
indirect detection experiments~\cite{DeRomeri:2022cem}.

\section{Summary and discussion}
\label{sec:conclusions}

The Scotogenic model is a very popular scenario for neutrino masses
and dark matter. In this work we have considered extensions of this
scenario that naturally explain the smallness of the quartic
$\lambda_5$ coupling and the origin of the Scotogenic \z2 parity. This
is achieved in UV extensions including a conserved global lepton
number symmetry, spontaneously broken by the VEVs of one or several
scalar singlets, and a new heavy state that suppresses all lepton
number violating effects at low energies. We explored all possible
models with these assumptions and found $50$ variations. They are all
characterized at low energies by the presence of a massless Goldstone
boson, the majoron, as well as other massive scalars besides the usual
Scotogenic states. Two specific example models are discussed in detail
in order to illustrate the basic ingredients of our setup. In these
two models, as well as in all the variants in our classification, a
rich phenomenology is expected, with potential signatures in collider
and lepton flavor violating searches, and implications for dark
matter.

Out of the $50$ models revealed by our analysis, only one had been
previously studied in the literature,
namely~\cite{Escribano:2021ymx}. This illustrates the vast model space
beyond the original Scotogenic model yet to be explored. In fact,
there are many variations of the fundamental setup that keep all the
positive features and include additional ingredients. While many of
these modified Scotogenic scenarios may contain unnecessary or
redundant ingredients, others may offer novel ways to address open
questions in current particle physics~\cite{Cepedello:2022xgb}. This
is the main motivation behind the classification presented in this
work.

There are several ways in which our analysis can be extended. First of
all, we have considered UV theories that realize the $\lambda_5$
coupling at tree-level. In this case, the only source of suppression
is given by the large energy scale $m_S$, assumed to lie well above
the electroweak scale. Alternatively, the $\lambda_5$ coupling can
also be realized at loop order, as recently explored
in~\cite{Abada:2022dvm}. This possibility leads to many novel
extensions of the Scotogenic setup with, at least potentially, new
phenomenological expectations. Another way in which our analysis can
be extended is by considering a local lepton number symmetry. In this
case, the massless majoron that was characteristic in our setup would
be replaced by a heavy $Z^\prime$ boson, with a dramatic impact on the
low-energy phenomenology. However, we note that this direction
requires non-trivial extensions of the fermion particle content in
order to cancel out the usual triangle gauge anomalies. Therefore, a
general classification of all possible gauge models becomes more
cumbersome, although interesting too. Finally, variations with
non-universal lepton charges for the $N$ fermions or featuring
alternative numbers of generations for the Scotogenic states can be
explored as well.

\section*{Acknowledgements}

The authors are grateful to Julio Leite for enlightening discussions,
in particular for drawing their attention to topology $\V$. Work
supported by the Spanish grants PID2020-113775GB-I00
(AEI/10.13039/501100011033) and CIPROM/2021/054 (Generalitat
Valenciana). The work of PE is supported by the FPI grant
PRE2018-084599.  AV acknowledges financial support from MINECO through
the Ramón y Cajal contract RYC2018-025795-I. DPS would like to thank
the AHEP group for the hospitality during his visit. The work of DPS
was supported by Ciencia de Frontera CONACYT project No.  428218 and
the program “BECAS CONACYT NACIONALES”.

\appendix

\section{Accidental \z2 symmetries}
\label{sec:app}

The dark \z2 parity of the Scotogenic model can also be an accidental
symmetry generated after the $\sigma$ singlet (or singlets) acquires a
VEV. In these scenarios, the symmetry breaking path is also $\rm
U(1)_L \to \mathbb{Z}_2$, but with $\ell_L$, $e_R$ and $\eta$ as the
only particles charged under the discrete symmetry. In this case, the
Yukawa term $\bar{N}\tilde{\eta}^{\dagger}\ell_L$ and the Majorana
mass $\overline{N}^c N$ are allowed by all symmetries, while
$\bar{N}\tilde{H}^{\dagger}\ell_L$ is forbidden. Furthermore, given
that $\eta$ is the only \z2-odd scalar, it will always appear in pairs
in the effective scalar potential. Therefore, although the \z2
Scotogenic parity does not emerge as a remnant symmetry after the
breaking of $\rm U(1)_L$, it appears \textit{accidentally} as a
consequence of it. In fact, one can see that the resulting symmetry is
nothing but a non-supersymmetric version of the well-known R-parity
$R_p = (-1)^{3B+L+2s}$~\cite{Farrar:1978xj}, which has its origin in
a combination of the $\rm U(1)_L$ and Lorentz
symmetries.~\footnote{The relation between R-parity and the Scotogenic
\z2 symmetry has been explored in~\cite{Kang:2019sab}.} These UV
models are not included in the classification presented in
Sec.~\ref{sec:clasif} since they violate requirement (A). However,
they also lead to the Scotogenic model at low energies.

{
\renewcommand{\arraystretch}{1.4}
\begin{table}[tb]
  \centering
  \begin{tabular}{|c|c||ccc|c|}
      \hline
      \textbf{Field} & \textbf{Generations} & \textbf{$\rm SU(3)_c$} & \textbf{$\rm SU(2)_L$} & \textbf{$\rm U(1)_Y$} & \textbf{$\rm U(1)_L$} \\
      \hline
      $\ell_L$ &  3  &  \one  &  \two  &  -1/2  &  1   \\
      $e_R$    &  3  &  \one  &  \one  &  -1    &  1   \\
      $N$      &  3  &  \one  &  \one  &  0     &  0   \\
      \hline
      $H$      &  1  &  \one  &  \two  &  1/2   &  0   \\
      $\eta$   &  1  &  \one  &  \two  &  1/2   &  -1  \\
      $\sigma$ &  1  &  \one  &  \one &  0     &  2   \\ 
      $S$      &  1  &  \one  &  \one  &  0     &  -1  \\
      \hline
  \end{tabular}
  \caption{\label{tab:ParticleContentAcc}
    Lepton and scalar particle content and representations under the gauge and global symmetries in an UV extension of the Scotogenic model with accidental \z2 symmetry.}
\end{table}
}

Let us illustrate this possibility with a specific
example.~\footnote{This model corresponds to the $\II^\prime\left(1,
\emptyset\right)$ model shown below in Table~\ref{tab:clasifAcc}.}
Consider the particle content and charge assignment in
Table~\ref{tab:ParticleContentAcc}. The new Yukawa interactions in the
model are given by
\begin{equation}
  \mathcal{L}_{\rm Y} = y \, \overline{N} \, \widetilde{\eta}^\dagger \, \ell_L + M_N \,  \overline{N}^c N + \hc \, ,
\end{equation}
while the scalar potential of the model is written as
\begin{equation}
\begin{split}
\mathcal{V}_{\rm UV} &= m_H^2 H^{\dagger} H+m_S^2 S^* S+m_{\sigma}^2 \sigma^* \sigma+ m_{\eta}^2\eta^{\dagger}\eta +\frac{\lambda_1}{2}(H^{\dagger}H)^2+\frac{\lambda_2}{2} (\eta^{\dagger} \eta)^2 \\
&+\frac{\lambda_S}{2}(S^* S)^2+\frac{\lambda_{\sigma}}{2}(\sigma^*\sigma)^2 + \lambda_3 (H^{\dagger}H)(\eta^{\dagger}\eta)+\lambda_3^S(H^{\dagger}H)(S^* S) \\
&+\lambda_3^{\sigma}(H^{\dagger}H)(\sigma^* \sigma)+\lambda_3^{\eta S}(\eta^{\dagger}\eta)(S^* S)+\lambda_3^{\eta \sigma}(\eta^{\dagger}\eta)(\sigma^*\sigma) \\
&+\lambda_3^{\sigma S}(\sigma^*\sigma)(S^* S)+\lambda_4 (H^{\dagger}\eta)(\eta^{\dagger}H)+\left[\beta (\sigma H^{\dagger}\eta S)+ \mu_1 \, H^\dagger \eta S^* + \mu_2 \, \sigma \, S^2 + \hc \right] \, .
\end{split}\label{UUVacc}
\end{equation}
It is easy to check that other Lagrangian terms are forbidden by $\rm
U(1)_L$. This global symmetry gets spontaneously broken once the
electroweak singlet $\sigma$ acquires a non-zero VEV, leaving a
remnant \z2 under which $\eta$, $S$, $\ell_L$ and $e_R$ are odd, while
the rest of the fields are even. We can call this symmetry
$\mathbb{Z}_2^{\rm rem}$. Since $q_N = 0$, $N$ is even under
$\mathbb{Z}_2^{\rm rem}$, and thus this symmetry cannot be identified
with the Scotogenic dark parity. Nevertheless, the Lagrangian of the
Scotogenic model is still obtained after decoupling the heavy scalar
$S$. This is due to the fact that a new accidental \z2 parity
appears. The only fields charged under this parity are $\eta$ and $N$,
while all the other fields in the effective theory are even,
therefore, this accidental symmetry, that we can denote as
$\mathbb{Z}_2^{\rm acc}$, is precisely the Scotogenic \z2. As already
explained, it is a non-supersymmetric version of R-parity.

Let us now generalize the idea studied in this Appendix. We consider
again the set of models in which $(H^{\dagger}\eta)^2$ is generated by
the topologies shown in Table~\ref{tab:topo} with the addition of at
most two different singlets $\sigma_{1,2}$. There are two
possibilities to construct models in which the $\mathbb{Z}_2^{\rm
  acc}$ symmetry is obtained:
\begin{enumerate}[label=(\roman*)]
\item \textbf{Models with $\boldsymbol{q_N \neq 0}$}. In this case we
  consider the models shown in Table~\ref{tab:topo} but impose that
  $N$ is even under the remnant $\mathbb{Z}_2^{\rm rem}$ parity while
  $\ell_L$, $e_R$ and $\eta$ are odd. The Majorana masses of the $N$
  fermions are induced by the $\kappa \, \sigma_1 \overline{N}^c N$
  Yukawa term.
\item \textbf{Models with $\boldsymbol{q_N = 0}$}. This case is
  excluded from the classification in Sec.~\ref{sec:clasif}, which
  focuses on $q_N \neq 0$, and must be discussed independently. In
  these models the Majorana mass term $M_N \, \overline{N}^c N$ is
  present in the UV theory.
\end{enumerate}
We now proceed to discuss these two cases independently. Again, we
find it convenient to consider topologies $\I-\IV$ and $\V$
separately, since they have some qualitative differences.

\subsection{Topologies I-IV}

We first consider topologies $\I-\IV$. The case of $q_N \neq 0$
can be regarded as a revision of our discussion in
Sec.~\ref{sec:clasif}, imposing now different conditions on the
resulting models. In fact, the models studied in Sec.~\ref{sec:clasif}
could also lead to $\rm U(1)_L \to \mathbb{Z}_2^{\rm rem}$, leaving
the Scotogenic \z2 parity as an accidental symmetry. This will be the
case when these conditions on $q_N$ are satisfied:
\begin{itemize}
\item $q_N = 2 \, z$, where $z$ can be any integer number except zero.
\item $q_N=\frac{\alpha}{\beta}$, with $\alpha, \beta \in \mathbb{Z}$
  and $\alpha$ and $\beta$ even and odd, respectively. Also,
  $\GCD(\alpha,\beta)=1$ has to be satisfied.
\end{itemize}
Notice, however, that models with fixed charges, that is, the ones
with only $\sigma_1$, always have the Scotogenic symmetry as the
remnant symmetry and do not enter this discussion.

Considering now scenarios with $q_N = 0$, only 11 different models
exist and they are listed in Table~\ref{tab:clasifAcc}. Let us denote
them as $\xi^{\prime}(A,B)$, where $\xi = \left\{ \I, \II, \III, \IV
\right\}$ and the prime is used to distinguish these models from the
ones studied in Sec.~\ref{sec:clasif}. Each of the 11 models needs to
satisfy any of the following conditions on $q_{\sigma_1}$ in order to
generate the \z2 parity as an accidental symmetry:
\begin{itemize}
\item $q_{\sigma_1} = 2 \, z$, where $z$ can be any integer number,
  including zero.~\footnote{We note that if $q_{\sigma_1} = 0$, a
  second $\sigma_2$ singlet, with $q_{\sigma_2} \neq 0$, is required
  to break the $\rm U(1)_L$ symmetry. In this case, $\sigma_1$ becomes
  a total singlet and is irrelevant for the model construction.}
\item $q_{\sigma_1} = \frac{\alpha}{\beta}$, with $\alpha, \beta \in
  \mathbb{Z}$ and $\alpha$ and $\beta$ even and odd,
  respectively. Also, $\GCD(\alpha,\beta)=1$ has to be satisfied.
\end{itemize}
We finally point out that in none of the above scenarios $\eta$ gets
an induced VEV.
\begin{table}[tp]
  \centering
  \renewcommand*{\arraystretch}{1.5}
\begin{tabular}{cccccccccc}\hline\hline
  & {\bf Topology} & $\boldsymbol{A}$ & $\boldsymbol{B}$ & $\boldsymbol{q_N}$ & $\boldsymbol{q_\eta}$ & $\boldsymbol{q_{\sigma_1}}$ & $\boldsymbol{q_{\sigma_2}}$ & $\boldsymbol{q_S}$ & $\boldsymbol{\left(\rm SU(2)_L, \rm U(1)_Y \right)_S}$ \\\hline\hline
  1 & $\I^\prime$ & $1$ & $\emptyset$ & $0$ & $-1$ & $2$ & - & $-2$ & $(\mathbf{3},1)$ \\
  2 & $\I^\prime$ & $\emptyset$ & $1$ & $0$ & $-1$ & $2$ & - & $0$ & $(\mathbf{3},1)$ \\
  3 & $\I^\prime$ & $1$ & $2$ & $0$ & $-1$ & $q_{\sigma_1}$ & $2-q_{\sigma_1}$ & $-q_{\sigma_1} $ & $(\mathbf{3},1)$ \\
  4-5 & $\II^\prime$ & $1$ & $\emptyset$ & $0$ & $-1$ & $2$ & - & $-1$ & $(\mathbf{3},0)$ or $(\mathbf{1},0)$ \\
  6-7 & $\II^\prime$ & $1$ & $2$ & $0$ & $-1$ & $q_{\sigma_1}$ & $2-q_{\sigma_1}$ & $1-q_{\sigma_1}$ & $(\mathbf{3},0)$ or $(\mathbf{1},0)$ \\
  8 & $\III^\prime$ & $1$ & $\emptyset$ & $0$ & $-1$ & $2$ & - & $-2$ & $(\mathbf{2},1/2)$ \\
  9 & $\III^\prime$ & $1$ & $2$ & $0$ & $-1$ & $q_{\sigma_1}$ & $2-q_{\sigma_1}$ & $-2$ & $(\mathbf{2},1/2)$ \\
  10 & $\IV^\prime$ & $1$ & $\emptyset$ & $0$ & $-1$ & $2$ & - & $1$ & $(\mathbf{2},1/2)$ \\
  11 & $\IV^\prime$ & $1$ & $2$ & $0$ & $-1$ & $q_{\sigma_1}$ & $2-q_{\sigma_1}$ & $1$ & $(\mathbf{2},1/2)$ \\
  \hline\hline
\end{tabular}
\caption{
UV extended models leading to topologies $\I-\IV$ and for which the
term $\overline{N}^c N$ is allowed and the Scotogenic \z2 is an
accidental symmetry. For each model we show the $\rm U(1)_L$ charges
of $N$, $\eta$, $\sigma_1$, $\sigma_2$ and $S$, as well as the $(\rm
SU(2)_L, \rm U(1)_Y)$ representation of $S$. Models that become any of
the models in this list after renaming the fields or redefining their
$\rm U(1)_L$ charges are not included. 
\label{tab:clasifAcc}
}
\end{table}

\subsection{Topology V}

We move on to topology $\V$. Again, for this topology we can
distinguish the same two types of models as for the previous
topologies. First of all, we consider the case $q_N \neq 0$. The
accidental symmetry arises for the models 29-40 in
Table~\ref{tab:clasifV} when any of these conditions is satisfied:
\begin{itemize}
\item $q_N = 2 \, z$, where $z$ can be any integer number except zero.
\item $q_N=\frac{\alpha}{\beta}$, with $\alpha, \beta \in \mathbb{Z}$
  and $\alpha$ and $\beta$ even and odd, respectively ($\beta \neq 1$). Also,
  $\GCD(\alpha,\beta)=1$ has to be satisfied.
\end{itemize}
For models 41-50 we have different conditions, although in all of them
we need $q_N=\frac{\alpha}{\beta}$, with $\alpha, \beta \in
\mathbb{Z}$ and $\alpha$ and $\beta$ even and odd, respectively. Also,
$\GCD(\alpha,\beta)=1$ has to be satisfied and we will allow $\beta = 1$ in these models. In addition:
\begin{itemize}
\item In model $\V(2,1,0)$, we further require $q_N \neq \pm \frac{2}{3}$ if $S$ is a singlet and $q_N \neq \frac{2}{3}$ if $S$ is a triplet. In both cases, $q_N$ can not be an integer (i.e., we need $\beta \neq 1$).
\item In model $\V(2,1^{*},0)$, we further require $q_N \neq \frac{2}{3}, \, \frac{2}{5}, \, \frac{2}{7}$ if $S$ is a singlet and $q_N \neq \frac{2}{3}, \, \frac{2}{5}$ if $S$ is a triplet. In both cases, $q_N$ can not be an integer (i.e., we need $\beta \neq 1$).
\item In model $\V(2,2,0)$, we have two options depending on the nature of $q_N$. If $q_N \in \mathbb{Z}$, $\GCD(3q_N, 1- q_N) = 1$ if $\frac{q_N -1}{3}$ is not an integer and $\GCD(q_N, \frac{q_N-1}{3}) = 1$ if $\frac{q_N -1}{3}$ is an integer. If $q_N \notin \mathbb{Z}$, $\GCD(3 \, \alpha,\alpha - \beta)= 1$ if $\frac{\alpha - \beta}{3}$ is not an integer and $\GCD(\alpha,\frac{\alpha - \beta}{3})= 1$ if $\frac{\alpha - \beta}{3}$ is an integer.
\item In model $\V(2,1^*,2)$, we also have two options depending on the nature of $q_N$. If $q_N \in \mathbb{Z}$, $\GCD(3q_N, 1- 2q_N) = 1$ if $\frac{1-2\,q_N}{3}$ is not an integer and $\GCD(q_N, \frac{1-2\,q_N}{3}) = 1$ if $\frac{1-2\,q_N}{3}$ is an integer. If $q_N \notin \mathbb{Z}$, $\GCD(3 \, \alpha,2\,\alpha - \beta)= 1$ if $\frac{2 \, \alpha - \beta}{3}$ is not an integer and $\GCD(\alpha,\frac{2\, \alpha - \beta}{3})= 1$ if $\frac{2\, \alpha - \beta}{3}$ is an integer.
\item In model $\V(2,1,2)$, we have again two options depending on the nature of $q_N$. There is no further requirement if $q_N \in \mathbb{Z}$, whereas if $q_N \notin \mathbb{Z}$, $\GCD(3 \, \alpha,\beta)= 1$ if $\frac{\beta}{3}$ is not an integer and $\GCD(\alpha, \frac{\beta}{3}) = 1$ if $\frac{\beta}{3}$ is an integer.
\end{itemize}
Models with $q_N = 0$ based on the topology $\V$ are collected in
Table~\ref{tab:clasifAccV}. Each of the 8 models needs to satisfy any
of the following conditions on $q_{\sigma_1}$ in order to generate the
\z2 parity as an accidental symmetry:
\begin{itemize}
\item $q_{\sigma_1} = 2 \, z$, where $z$ can be any integer number,
  including zero.
\item $q_{\sigma_1} = \frac{\alpha}{\beta}$, with $\alpha, \beta \in
  \mathbb{Z}$ and $\alpha$ and $\beta$ even and odd,
  respectively. Also, $\GCD(\alpha,\beta)=1$ 
has to be satisfied.
\end{itemize}

\begin{table}[tp]
  \centering
  \renewcommand*{\arraystretch}{1.5}
\begin{tabular}{ccccccccccc}\hline\hline
  & {\bf Topology} & $\boldsymbol{A}$ & $\boldsymbol{B}$ & $\boldsymbol{C}$ & $\boldsymbol{q_N}$ & $\boldsymbol{q_\eta}$ & $\boldsymbol{q_{\sigma_1}}$ & $\boldsymbol{q_{\sigma_2}}$ & $\boldsymbol{q_S}$ & $\boldsymbol{\left(\rm SU(2)_L, \rm U(1)_Y \right)_S}$ \\\hline\hline
  12-13 & $\V^\prime$ & $1$ & $1$ & $\emptyset$ & $0$ & $-1$ & $\frac{2}{3}$ & - & $\frac{1}{3}$ & $(\mathbf{3},0)$ or $(\mathbf{1},0)$ \\
  14-15 & $\V^\prime$ & $\emptyset$ & $1$ & $2$ & $0$ & $-1$ & $q_{\sigma_1}$ & $2-q_{\sigma_1}$ & $1$ & $(\mathbf{3},0)$ or $(\mathbf{1},0)$ \\
  16-17 & $\V^\prime$ & $1$ & $2$ & $\emptyset$ & $0$ & $-1$ & $q_{\sigma_1}$ & $2-2q_{\sigma_1}$ & $1-q_{\sigma_1}$ & $(\mathbf{3},0)$ or $(\mathbf{1},0)$ \\
  18-19 & $\V^\prime$ & $1$ & $1$ & $2$ & $0$ & $-1$ & $q_{\sigma_1}$ & $2-3q_{\sigma_1}$ & $1-q_{\sigma_1}$ & $(\mathbf{3},0)$ or $(\mathbf{1},0)$ \\
  \hline\hline
\end{tabular}
\caption{
UV extended models leading to topology $\V$ and for which the term
$\overline{N}^c N$ is allowed and the Scotogenic \z2 is an accidental
symmetry. For each model we show the $\rm U(1)_L$ charges of $N$,
$\eta$, $\sigma_1$, $\sigma_2$ and $S$, as well as the $(\rm SU(2)_L,
\rm U(1)_Y)$ representation of $S$. Models that become any of the
models in this list after renaming the fields or redefining their $\rm
U(1)_L$ charges are not included.
\label{tab:clasifAccV}}
\end{table}

\bibliographystyle{utphys}
\bibliography{refs}

\providecommand{\href}[2]{#2}\begingroup\raggedright\begin{thebibliography}{10}

\bibitem{Ma:2006km}
E.~Ma, ``{Verifiable radiative seesaw mechanism of neutrino mass and dark
  matter},'' \href{http://dx.doi.org/10.1103/PhysRevD.73.077301}{{\em Phys.
  Rev.} {\bfseries D73} (2006) 077301},
\href{http://arxiv.org/abs/hep-ph/0601225}{{\ttfamily arXiv:hep-ph/0601225
  [hep-ph]}}.

\bibitem{Zee:1980ai}
A.~Zee, ``{A Theory of Lepton Number Violation, Neutrino Majorana Mass, and
  Oscillation},'' \href{http://dx.doi.org/10.1016/0370-2693(80)90349-4,
  10.1016/0370-2693(80)90193-8}{{\em Phys. Lett.} {\bfseries 93B} (1980) 389}.
[Erratum: Phys. Lett.95B,461(1980)].

\bibitem{Cheng:1980qt}
T.~P. Cheng and L.-F. Li, ``{Neutrino Masses, Mixings and Oscillations in $\rm
  SU(2) \times U(1)$ Models of Electroweak Interactions},''
\href{http://dx.doi.org/10.1103/PhysRevD.22.2860}{{\em Phys. Rev.} {\bfseries
  D22} (1980) 2860}.

\bibitem{Zee:1985id}
A.~Zee, ``{Quantum Numbers of Majorana Neutrino Masses},''
\href{http://dx.doi.org/10.1016/0550-3213(86)90475-X}{{\em Nucl. Phys.}
  {\bfseries B264} (1986) 99--110}.

\bibitem{Babu:1988ki}
K.~S. Babu, ``{Model of 'Calculable' Majorana Neutrino Masses},''
\href{http://dx.doi.org/10.1016/0370-2693(88)91584-5}{{\em Phys. Lett.}
  {\bfseries B203} (1988) 132--136}.

\bibitem{Cai:2017jrq}
Y.~Cai, J.~Herrero-Garc\'\i{}a, M.~A. Schmidt, A.~Vicente, and R.~R. Volkas,
  ``{From the trees to the forest: a review of radiative neutrino mass
  models},'' \href{http://dx.doi.org/10.3389/fphy.2017.00063}{{\em Front. in
  Phys.} {\bfseries 5} (2017) 63},
  \href{http://arxiv.org/abs/1706.08524}{{\ttfamily arXiv:1706.08524
  [hep-ph]}}.

\bibitem{tHooft:1979rat}
G.~'t~Hooft, ``{Naturalness, chiral symmetry, and spontaneous chiral symmetry
  breaking},'' \href{http://dx.doi.org/10.1007/978-1-4684-7571-5_9}{{\em NATO
  Sci. Ser. B} {\bfseries 59} (1980) 135--157}.

\bibitem{Escribano:2021ymx}
P.~Escribano and A.~Vicente, ``{An ultraviolet completion for the Scotogenic
  model},'' \href{http://dx.doi.org/10.1016/j.physletb.2021.136717}{{\em Phys.
  Lett. B} {\bfseries 823} (2021) 136717},
  \href{http://arxiv.org/abs/2107.10265}{{\ttfamily arXiv:2107.10265
  [hep-ph]}}.

\bibitem{Chikashige:1980ui}
Y.~Chikashige, R.~N. Mohapatra, and R.~D. Peccei, ``{Are There Real Goldstone
  Bosons Associated with Broken Lepton Number?},''
  \href{http://dx.doi.org/10.1016/0370-2693(81)90011-3}{{\em Phys. Lett. B}
  {\bfseries 98} (1981) 265--268}.

\bibitem{Gelmini:1980re}
G.~Gelmini and M.~Roncadelli, ``{Left-Handed Neutrino Mass Scale and
  Spontaneously Broken Lepton Number},''
  \href{http://dx.doi.org/10.1016/0370-2693(81)90559-1}{{\em Phys. Lett. B}
  {\bfseries 99} (1981) 411--415}.

\bibitem{Schechter:1981cv}
J.~Schechter and J.~W.~F. Valle, ``{Neutrino Decay and Spontaneous Violation of
  Lepton Number},'' \href{http://dx.doi.org/10.1103/PhysRevD.25.774}{{\em Phys.
  Rev. D} {\bfseries 25} (1982) 774}.

\bibitem{Aulakh:1982yn}
C.~Aulakh and R.~N. Mohapatra, ``{Neutrino as the Supersymmetric Partner of the
  Majoron},'' \href{http://dx.doi.org/10.1016/0370-2693(82)90262-3}{{\em Phys.
  Lett. B} {\bfseries 119} (1982) 136--140}.

\bibitem{Escribano:2020iqq}
P.~Escribano, M.~Reig, and A.~Vicente, ``{Generalizing the Scotogenic model},''
  \href{http://dx.doi.org/10.1007/JHEP07(2020)097}{{\em JHEP} {\bfseries 07}
  (2020) 097}, \href{http://arxiv.org/abs/2004.05172}{{\ttfamily
  arXiv:2004.05172 [hep-ph]}}.

\bibitem{Sun:2021jpw}
J.~Sun, Y.~Cheng, and X.-G. He, ``{Structure Of Flavor Changing Goldstone Boson
  Interactions},'' \href{http://dx.doi.org/10.1007/JHEP04(2021)141}{{\em JHEP}
  {\bfseries 04} (2021) 141}, \href{http://arxiv.org/abs/2101.06055}{{\ttfamily
  arXiv:2101.06055 [hep-ph]}}.

\bibitem{Escribano:2020wua}
P.~Escribano and A.~Vicente, ``{Ultralight scalars in leptonic observables},''
  \href{http://dx.doi.org/10.1007/JHEP03(2021)240}{{\em JHEP} {\bfseries 03}
  (2021) 240}, \href{http://arxiv.org/abs/2008.01099}{{\ttfamily
  arXiv:2008.01099 [hep-ph]}}.

\bibitem{Calibbi:2020jvd}
L.~Calibbi, D.~Redigolo, R.~Ziegler, and J.~Zupan, ``{Looking forward to
  lepton-flavor-violating ALPs},''
  \href{http://dx.doi.org/10.1007/JHEP09(2021)173}{{\em JHEP} {\bfseries 09}
  (2021) 173}, \href{http://arxiv.org/abs/2006.04795}{{\ttfamily
  arXiv:2006.04795 [hep-ph]}}.

\bibitem{Croon:2020lrf}
D.~Croon, G.~Elor, R.~K. Leane, and S.~D. McDermott, ``{Supernova Muons: New
  Constraints on $Z$' Bosons, Axions and ALPs},''
  \href{http://dx.doi.org/10.1007/JHEP01(2021)107}{{\em JHEP} {\bfseries 01}
  (2021) 107}, \href{http://arxiv.org/abs/2006.13942}{{\ttfamily
  arXiv:2006.13942 [hep-ph]}}.

\bibitem{Hirsch:2009ee}
M.~Hirsch, A.~Vicente, J.~Meyer, and W.~Porod, ``{Majoron emission in muon and
  tau decays revisited},''
  \href{http://dx.doi.org/10.1103/PhysRevD.79.055023}{{\em Phys. Rev. D}
  {\bfseries 79} (2009) 055023},
  \href{http://arxiv.org/abs/0902.0525}{{\ttfamily arXiv:0902.0525 [hep-ph]}}.
  [Erratum: Phys.Rev.D 79, 079901 (2009)].

\bibitem{Escribano:2021uhf}
P.~Escribano, M.~Hirsch, J.~Nava, and A.~Vicente, ``{Observable flavor
  violation from spontaneous lepton number breaking},''
  \href{http://dx.doi.org/10.1007/JHEP01(2022)098}{{\em JHEP} {\bfseries 01}
  (2022) 098}, \href{http://arxiv.org/abs/2108.01101}{{\ttfamily
  arXiv:2108.01101 [hep-ph]}}.

\bibitem{Pilaftsis:1993af}
A.~Pilaftsis, ``{Astrophysical and terrestrial constraints on singlet Majoron
  models},'' \href{http://dx.doi.org/10.1103/PhysRevD.49.2398}{{\em Phys. Rev.
  D} {\bfseries 49} (1994) 2398--2404},
  \href{http://arxiv.org/abs/hep-ph/9308258}{{\ttfamily arXiv:hep-ph/9308258}}.

\bibitem{Heeck:2019guh}
J.~Heeck and H.~H. Patel, ``{Majoron at two loops},''
  \href{http://dx.doi.org/10.1103/PhysRevD.100.095015}{{\em Phys. Rev. D}
  {\bfseries 100} no.~9, (2019) 095015},
  \href{http://arxiv.org/abs/1909.02029}{{\ttfamily arXiv:1909.02029
  [hep-ph]}}.

\bibitem{Babu:2007sm}
K.~S. Babu and E.~Ma, ``{Singlet fermion dark matter and electroweak
  baryogenesis with radiative neutrino mass},''
  \href{http://dx.doi.org/10.1142/S0217751X08040299}{{\em Int. J. Mod. Phys. A}
  {\bfseries 23} (2008) 1813--1819},
  \href{http://arxiv.org/abs/0708.3790}{{\ttfamily arXiv:0708.3790 [hep-ph]}}.

\bibitem{Casas:2001sr}
J.~A. Casas and A.~Ibarra, ``{Oscillating neutrinos and $\mu \to e, \gamma$},''
  \href{http://dx.doi.org/10.1016/S0550-3213(01)00475-8}{{\em Nucl. Phys. B}
  {\bfseries 618} (2001) 171--204},
  \href{http://arxiv.org/abs/hep-ph/0103065}{{\ttfamily arXiv:hep-ph/0103065}}.

\bibitem{Toma:2013zsa}
T.~Toma and A.~Vicente, ``{Lepton Flavor Violation in the Scotogenic Model},''
  \href{http://dx.doi.org/10.1007/JHEP01(2014)160}{{\em JHEP} {\bfseries 01}
  (2014) 160}, \href{http://arxiv.org/abs/1312.2840}{{\ttfamily arXiv:1312.2840
  [hep-ph]}}.

\bibitem{Cordero-Carrion:2018xre}
I.~Cordero-Carri\'on, M.~Hirsch, and A.~Vicente, ``{Master Majorana neutrino
  mass parametrization},''
  \href{http://dx.doi.org/10.1103/PhysRevD.99.075019}{{\em Phys. Rev. D}
  {\bfseries 99} no.~7, (2019) 075019},
  \href{http://arxiv.org/abs/1812.03896}{{\ttfamily arXiv:1812.03896
  [hep-ph]}}.

\bibitem{Cordero-Carrion:2019qtu}
I.~Cordero-Carri\'on, M.~Hirsch, and A.~Vicente, ``{General parametrization of
  Majorana neutrino mass models},''
  \href{http://dx.doi.org/10.1103/PhysRevD.101.075032}{{\em Phys. Rev. D}
  {\bfseries 101} no.~7, (2020) 075032},
  \href{http://arxiv.org/abs/1912.08858}{{\ttfamily arXiv:1912.08858
  [hep-ph]}}.

\bibitem{deSalas:2020pgw}
P.~F. de~Salas, D.~V. Forero, S.~Gariazzo, P.~Mart\'\i{}nez-Mirav\'e, O.~Mena,
  C.~A. Ternes, M.~T\'ortola, and J.~W.~F. Valle, ``{2020 global reassessment
  of the neutrino oscillation picture},''
  \href{http://dx.doi.org/10.1007/JHEP02(2021)071}{{\em JHEP} {\bfseries 02}
  (2021) 071}, \href{http://arxiv.org/abs/2006.11237}{{\ttfamily
  arXiv:2006.11237 [hep-ph]}}.

\bibitem{LHCHiggsCrossSectionWorkingGroup:2016ypw}
{\bfseries LHC Higgs Cross Section Working Group} Collaboration, D.~de~Florian
  {\em et~al.}, ``{Handbook of LHC Higgs Cross Sections: 4. Deciphering the
  Nature of the Higgs Sector},''
  \href{http://arxiv.org/abs/1610.07922}{{\ttfamily arXiv:1610.07922
  [hep-ph]}}.

\bibitem{CMS:2018yfx}
{\bfseries CMS} Collaboration, A.~M. Sirunyan {\em et~al.}, ``{Search for
  invisible decays of a Higgs boson produced through vector boson fusion in
  proton-proton collisions at $\sqrt{s} =$ 13 TeV},''
  \href{http://dx.doi.org/10.1016/j.physletb.2019.04.025}{{\em Phys. Lett. B}
  {\bfseries 793} (2019) 520--551},
  \href{http://arxiv.org/abs/1809.05937}{{\ttfamily arXiv:1809.05937
  [hep-ex]}}.

\bibitem{Biekotter:2022ckj}
T.~Biek\"otter and M.~Pierre, ``{Higgs-boson visible and invisible constraints
  on hidden sectors},''
  \href{http://dx.doi.org/10.1140/epjc/s10052-022-10990-x}{{\em Eur. Phys. J.
  C} {\bfseries 82} no.~11, (2022) 1026},
  \href{http://arxiv.org/abs/2208.05505}{{\ttfamily arXiv:2208.05505
  [hep-ph]}}.

\bibitem{Deshpande:1977rw}
N.~G. Deshpande and E.~Ma, ``{Pattern of Symmetry Breaking with Two Higgs
  Doublets},'' \href{http://dx.doi.org/10.1103/PhysRevD.18.2574}{{\em Phys.
  Rev. D} {\bfseries 18} (1978) 2574}.

\bibitem{Barbieri:2006dq}
R.~Barbieri, L.~J. Hall, and V.~S. Rychkov, ``{Improved naturalness with a
  heavy Higgs: An Alternative road to LHC physics},''
  \href{http://dx.doi.org/10.1103/PhysRevD.74.015007}{{\em Phys. Rev. D}
  {\bfseries 74} (2006) 015007},
  \href{http://arxiv.org/abs/hep-ph/0603188}{{\ttfamily arXiv:hep-ph/0603188}}.

\bibitem{LopezHonorez:2006gr}
L.~Lopez~Honorez, E.~Nezri, J.~F. Oliver, and M.~H.~G. Tytgat, ``{The Inert
  Doublet Model: An Archetype for Dark Matter},''
  \href{http://dx.doi.org/10.1088/1475-7516/2007/02/028}{{\em JCAP} {\bfseries
  02} (2007) 028}, \href{http://arxiv.org/abs/hep-ph/0612275}{{\ttfamily
  arXiv:hep-ph/0612275}}.

\bibitem{LopezHonorez:2010eeh}
L.~Lopez~Honorez and C.~E. Yaguna, ``{The inert doublet model of dark matter
  revisited},'' \href{http://dx.doi.org/10.1007/JHEP09(2010)046}{{\em JHEP}
  {\bfseries 09} (2010) 046}, \href{http://arxiv.org/abs/1003.3125}{{\ttfamily
  arXiv:1003.3125 [hep-ph]}}.

\bibitem{Diaz:2015pyv}
M.~A. D\'\i{}az, B.~Koch, and S.~Urrutia-Quiroga, ``{Constraints to Dark Matter
  from Inert Higgs Doublet Model},''
  \href{http://dx.doi.org/10.1155/2016/8278375}{{\em Adv. High Energy Phys.}
  {\bfseries 2016} (2016) 8278375},
  \href{http://arxiv.org/abs/1511.04429}{{\ttfamily arXiv:1511.04429
  [hep-ph]}}.

\bibitem{Kubo:2006yx}
J.~Kubo, E.~Ma, and D.~Suematsu, ``{Cold Dark Matter, Radiative Neutrino Mass,
  $\mu \to e\gamma$, and Neutrinoless Double Beta Decay},''
  \href{http://dx.doi.org/10.1016/j.physletb.2006.08.085}{{\em Phys. Lett. B}
  {\bfseries 642} (2006) 18--23},
  \href{http://arxiv.org/abs/hep-ph/0604114}{{\ttfamily arXiv:hep-ph/0604114}}.

\bibitem{AristizabalSierra:2008cnr}
D.~Aristizabal~Sierra, J.~Kubo, D.~Restrepo, D.~Suematsu, and O.~Zapata,
  ``{Radiative seesaw: Warm dark matter, collider and lepton flavour violating
  signals},'' \href{http://dx.doi.org/10.1103/PhysRevD.79.013011}{{\em Phys.
  Rev. D} {\bfseries 79} (2009) 013011},
  \href{http://arxiv.org/abs/0808.3340}{{\ttfamily arXiv:0808.3340 [hep-ph]}}.

\bibitem{Suematsu:2009ww}
D.~Suematsu, T.~Toma, and T.~Yoshida, ``{Reconciliation of CDM abundance and
  $\mu \to e \gamma$ in a radiative seesaw model},''
  \href{http://dx.doi.org/10.1103/PhysRevD.79.093004}{{\em Phys. Rev. D}
  {\bfseries 79} (2009) 093004},
  \href{http://arxiv.org/abs/0903.0287}{{\ttfamily arXiv:0903.0287 [hep-ph]}}.

\bibitem{Adulpravitchai:2009gi}
A.~Adulpravitchai, M.~Lindner, and A.~Merle, ``{Confronting Flavour Symmetries
  and extended Scalar Sectors with Lepton Flavour Violation Bounds},''
  \href{http://dx.doi.org/10.1103/PhysRevD.80.055031}{{\em Phys. Rev. D}
  {\bfseries 80} (2009) 055031},
  \href{http://arxiv.org/abs/0907.2147}{{\ttfamily arXiv:0907.2147 [hep-ph]}}.

\bibitem{Vicente:2014wga}
A.~Vicente and C.~E. Yaguna, ``{Probing the scotogenic model with lepton flavor
  violating processes},'' \href{http://dx.doi.org/10.1007/JHEP02(2015)144}{{\em
  JHEP} {\bfseries 02} (2015) 144},
  \href{http://arxiv.org/abs/1412.2545}{{\ttfamily arXiv:1412.2545 [hep-ph]}}.

\bibitem{Bonilla:2019ipe}
C.~Bonilla, L.~M.~G. de~la Vega, J.~M. Lamprea, R.~A. Lineros, and E.~Peinado,
  ``{Fermion Dark Matter and Radiative Neutrino Masses from Spontaneous Lepton
  Number Breaking},'' \href{http://dx.doi.org/10.1088/1367-2630/ab7254}{{\em
  New J. Phys.} {\bfseries 22} no.~3, (2020) 033009},
  \href{http://arxiv.org/abs/1908.04276}{{\ttfamily arXiv:1908.04276
  [hep-ph]}}.

\bibitem{DeRomeri:2022cem}
V.~De~Romeri, J.~Nava, M.~Puerta, and A.~Vicente, ``{Dark matter in the
  Scotogenic model with spontaneous lepton number violation},''
  \href{http://arxiv.org/abs/2210.07706}{{\ttfamily arXiv:2210.07706
  [hep-ph]}}.

\bibitem{Cepedello:2022xgb}
R.~Cepedello, P.~Escribano, and A.~Vicente, ``{Neutrino masses, flavor
  anomalies and muon $\boldsymbol{g-2}$ from dark loops},''
  \href{http://arxiv.org/abs/2209.02730}{{\ttfamily arXiv:2209.02730
  [hep-ph]}}.

\bibitem{Abada:2022dvm}
A.~Abada, N.~Bernal, A.~E. C\'arcamo~Hern\'andez, S.~Kovalenko, T.~B. de~Melo,
  and T.~Toma, ``{Phenomenological and cosmological implications of a
  scotogenic three-loop neutrino mass model},''
  \href{http://arxiv.org/abs/2212.06852}{{\ttfamily arXiv:2212.06852
  [hep-ph]}}.

\bibitem{Farrar:1978xj}
G.~R. Farrar and P.~Fayet, ``{Phenomenology of the Production, Decay, and
  Detection of New Hadronic States Associated with Supersymmetry},''
  \href{http://dx.doi.org/10.1016/0370-2693(78)90858-4}{{\em Phys. Lett. B}
  {\bfseries 76} (1978) 575--579}.

\bibitem{Kang:2019sab}
S.~K. Kang, O.~Popov, R.~Srivastava, J.~W.~F. Valle, and C.~A. Vaquera-Araujo,
  ``{Scotogenic dark matter stability from gauged matter parity},''
  \href{http://dx.doi.org/10.1016/j.physletb.2019.135013}{{\em Phys. Lett. B}
  {\bfseries 798} (2019) 135013},
  \href{http://arxiv.org/abs/1902.05966}{{\ttfamily arXiv:1902.05966
  [hep-ph]}}.

\end{thebibliography}\endgroup

\end{document}